\documentclass[11pt,a4paper]{article}
\usepackage[T1]{fontenc}
\usepackage[utf8]{inputenc}
\usepackage{lmodern}
\usepackage[margin=2.5cm]{geometry}
\usepackage{graphicx}
\usepackage{pdflscape}
\usepackage{eso-pic}
\newcommand{\landscapepagenumber}{\thispagestyle{empty}%
  \AddToShipoutPictureFG*{\AtPageCenter{\put(\LenToUnit{\dimexpr0.5\paperwidth-0.45in\relax},0){\rotatebox{90}{\makebox(0,0){\thepage}}}}}}
\usepackage{amsmath}
\usepackage{microtype}
\usepackage{xcolor}
\usepackage{booktabs}
\usepackage{caption}
\newcommand{\capFigOne}{%
\textbf{The brain-and-nerve-cord connectome as a spiking network inside a fly body.}
\textbf{a}, The body model from above and from the side, with the connectome inside it at physical scale.
\textbf{b}, The neurons of the connectome drawn as their arbors and colored by region, from above and from the side.
\textbf{c}, Synapse counts between the 24 regions as a chord diagram; each ribbon joins a presynaptic region to a postsynaptic region and its width is the number of synapses between them.
\textbf{d}, The 24 regions used throughout, one tile each; the region's neurons are colored on the gray outline of the whole central nervous system.
\textbf{e}, The synapse counts of \textbf{c} as a matrix, presynaptic region by row and postsynaptic region by column, on a logarithmic color scale (1 to 9.2 million synapses); rows and columns are ordered by hierarchical clustering, shown by the dendrograms; the colored strips give each region's color.
\textbf{f}, Functional influence: each region was stimulated on its own, at 40 Hz for 3 s in the network outside the body, and the spiking evoked in every region was recorded. Rows, the stimulated region; columns, the responding region, in the order of \textbf{e}; each row is normalized to its maximum (0 to 1); mean over three seeds.
The body model is NeuroMechFly\cite{WangChen2024} and the connectome is BANC\cite{Bates2026}.%
}

\newcommand{\capFigTwo}{%
\textbf{Identified pathways driven in the network outside the body and read at identified target neurons.}
\textbf{a}, The six behavior families on an outline of the fly: the driven population (colored), the readout, and in italics the behavior in the animal.
\textbf{b}, Every driven population on the central nervous system, with its neuron count.
\textbf{c}, Every relay neuron between the sugar gustatory receptor neurons (GRNs) and the proboscis motor neuron MN9 (columns), its firing rate (color) under sugar at 10 to 200 Hz and under sugar at 200 Hz with a second tastant (bitter, Ir94e, or water as the control); above, each relay's synapses onto MN9 (bars, square-root scale) and the share of its input from the sugar GRNs (dots; dotted line, the band of \textbf{f}).
\textbf{d}, As \textbf{c} for escape: the neurons feeding the giant fiber, under LC4, LPLC2, both, LPLC1, or LC11 as the control.
\textbf{e}, Every leg motor neuron (columns by leg; each group's label gives how many respond) under each descending drive (rows), intact and rewired; bars above, its descending input (square-root scale).
\textbf{f}, The target neuron's firing rate against the share of its total input that the driven pathway supplies, on logarithmic axes: filled circles, escape pathways onto the giant fiber; diamond, taste onto MN9; open circles, the giant fiber onto TTMn with a gap junction added by hand at several weights; square, seven pathways at or below 0.03\% of input, named; shading, the operating range.
\textbf{g}, MN9's rate under sugar alone and with a second tastant, intact (green) and with the tastant's outgoing synapses shuffled (blue); bars, mean; caps, s.e.m.; dots, nine seeds; dotted, the unsuppressed baseline.
\textbf{h}, Each pathway's registered outcome: conducts, dies under the null (the shuffle), meets the criterion registered for the animal's behavior (filled, yes; open, no; half, conditionally), and the reason where not.
\textbf{i}, The driven side's rate above control against the network's background rate (0 to 20 Hz): filled, mean $\pm$ s.e.m. over nine seeds; open, the other side; crosses, the rewired twin; diamond, MN9; shading, the rates at which the pathway conducts.%
}

\newcommand{\capFigThree}{%
\textbf{The antennal-lobe circuit and the Kenyon-cell odor code, as read and with the cholinergic gain scaled.}
Blue, as read; vermillion, the matched variant (the cholinergic local neurons' output scaled to 0.15); green, the fly's window; all outside the body.
\textbf{a}, The circuit: cell counts in ovals, synapse counts on arrows (arrowhead excitatory, bar inhibitory), local neurons by predicted transmitter, the scaled output in both colors.
\textbf{b}, Each local neuron's output synapses (logarithmic) against its prediction score, colored by class.
\textbf{c}, Active fraction (dashed) and the two odors' top-5\% overlap (solid) per layer: olfactory receptor neurons (ORN), projection neurons (PN), local neurons (LN), Kenyon cells (KC), lateral-horn output neurons (LHON) and mushroom-body output neurons (MBON); six seeds each.
\textbf{d}, Kenyon-cell (solid) and projection-neuron (dashed) active fraction and overlap against the cholinergic gain (1 = as read); six seeds; shading, the matched range.
\textbf{e}, Kenyon-cell active fraction at the matched gain for eight odors of five glomeruli (labeled by first and last glomerulus); lines join as-read and matched; three or six seeds.
\textbf{f}, MBON spikes per presentation against the gain, odor A filled and B open, and per odor at the matched gain (bars) with the as-read range (band); dotted, 5 spikes.
\textbf{g}, Kenyon-cell active fraction with every receptor driven at one rate and no odor (filled, after a warm-up; hollow, from silence); right, against the glomeruli per odor; six and three seeds.
\textbf{h}, After training on odor A: MBON depression (fraction of before) of A (filled) and B (open) against the glomeruli B shares with A; right, B's as a fraction of A's at training strengths 1 and 0.5 (matched) and 0.5 (as read, from the median depressions), Kenyon-cell overlap in parentheses; three seeds.
\textbf{i}, One cholinergic type silenced at a time: the fall in Kenyon-cell active fraction (bars) and the type's output synapses (logarithmic); Spearman $\rho=0.78$, 11 types, two-sided $P=0.0045$; three seeds.
\textbf{j}, Types silenced together (filled) against the sum of their single falls (open); dotted, all 105.
\textbf{k}, Kenyon-cell active fraction, overlap and MBON spikes against the odor drive (fraction of full); six seeds at drive 1, three below.%
}

\newcommand{\capFigFour}{%
\textbf{The embodied fly in nine arenas under three wiring conditions.}
Intact (blue), rewired (vermillion) and mushroom-body-lesioned (pink); 20 seeds (simulated flies with their own initial conditions) per condition in each of nine arenas ordered by clutter, i to ix, with 32 odor sources in all: 180 runs per condition.
\textbf{a}, Every seed's path in every arena, the three conditions overlaid; green dots, the sources; gray bars, the walls.
\textbf{b}--\textbf{d}, Occupancy per condition, visits per bin on a shared logarithmic color scale; the label at each source is the fraction of seeds that reached it.
\textbf{e}, Firing measured on the central nervous system in one run per condition (arena i, seed 0), spikes per neuron per step on a logarithmic color scale; the olfactory region, mushroom body and lateral horn are labeled.
\textbf{f}, Odor sources reached per seed, of 32: dots, seeds; bars, means; dashed lines, all 32 and the chance level of a correlated random walk at the fly's own speed.
\textbf{g}, The fraction of sources reached, by arena: large dots, means; small dots, seeds.
\textbf{h}, The fraction of seeds reaching each source in order along the corridor.
\textbf{i}--\textbf{k}, Path straightness (net displacement over path length), median walking speed and the fraction of time moving: dots, runs; bars, means.
\textbf{l}, \textbf{m}, Measured firing rate and the fraction of neurons firing, by structure, from the same single run per condition as \textbf{e}; open markers, zero.%
}

\newcommand{\capFigFive}{%
\textbf{Perturbations of the sensorimotor loop in the body.}
\textbf{a}, The loop drawn on the body, with the six manipulations numbered as in the key.
\textbf{b}, Paths in the forage arena (three sources, green) by wiring condition; $n=20$ seeds each; bar, 10 mm.
\textbf{c}, Odor sources reached (of 3) per seed by condition: dots, seeds; bars, means; above each, the mean's difference from intact; dotted, the intact mean; $n=20$.
\textbf{d}, Six measures for the same conditions, mean with s.e.m. below; color, the value as a fraction of intact (abolished to spared).
\textbf{e}, Closed loop (blue) against motor replay (yellow), the closed-loop run's own motor commands fed back with the sensory loop cut, both under a 30$^{\circ}$ yaw perturbation: two arenas with every seed's path; right, each run's two outcomes over nine arenas; dashed, the diagonal.
\textbf{f}, Net turn and speed change over 2 s of drive at each descending-neuron target, intact (blue) and rewired (vermillion): dots, seeds; bars, means; gray band, no-drive mean $\pm$ s.e.m.; dotted, zero; the marker above each target, the sign measured in the animal (filled, carried; open, not); 20 seeds per target intact, 7 per shuffle rewired.
\textbf{g}, The fraction of sources reached per run under closed loop (blue) and replay (yellow) by arena and over all nine; small dots, the unperturbed closed-loop record; bars, means. Two-sided paired Wilcoxon signed-rank test on the 20 flies' mean fractions: $W=0$, exact $P=1.9\times10^{-6}$; all 180 pairs and all nine arenas fall the same way.
\textbf{h}, \textbf{i}, Navigation (sources reached, closest approach) and walking (leg motor rate, path length) against the fraction of neurons removed at random (green), by traffic (black) or by traffic within the brain (dashed): lines, mean; shading, s.e.m. over 9 to 20 runs; small dots, seeds; dotted, intact; \textbf{i}'s top strip, the fraction of runs excluded for a physics fault.
\textbf{j}, The sign of each driven behavior in the wiring alone, silent and on a 2 Hz baseline (Fig.~2), and in the body at three receptor rates: green, the sign measured in the animal; gray, the opposite; open, no effect (Methods); that sign under each column.%
}

\newcommand{\capFigSix}{%
\textbf{Dopamine-gated plasticity at the Kenyon-cell-to-MBON synapses, run inside the loop.}
\textbf{a}, The six elements of the learning, numbered on the body and named beside it. Right, the protocol: six training episodes between a naive and a post test, each an odor bar over a dopamine bar, under odor + dopamine (paired, vermillion), dopamine alone (unpaired, blue) or no plasticity ($\eta=0$, gray).
\textbf{b}, Left, the plastic connections' mean weight over training by condition ($n=20$); right, the paired weight after training against the learning rate $\eta$, every seed; dotted, the naive weight.
\textbf{c}, Kenyon cells firing per episode with the odor (paired) and without it (dopamine alone), both models, one seed per condition; percentages, means over training (shaded).
\textbf{d}, Spikes per step of three MBON groups in the naive test (dark blue) and after training, paired and dopamine alone, both models: the 41 approach-driving MBONs of the aversive compartments (plastic targets), the 34 other approach-driving MBONs and the 25 avoidance-driving MBONs.
\textbf{e}, Top, MBON valence before and after training, every seed; dotted, the 0.7 that rescues approach (Fig.~5); gray line, zero (below it the fly turns away). Bottom, the valence through the protocol, one seed per condition.
\textbf{f}, Closest approach after training by condition: dots, seeds; bars, means; dotted, the capture radius (4 mm) and the naive value. Two-sided Mann--Whitney $U$ tests with Cliff's $\delta$, $n=20$ per group: paired and unpaired against no plasticity, each $P=6.8\times10^{-8}$, $\delta=+1.00$; paired against unpaired $P=0.49$, $\delta=+0.13$.
\textbf{g}, Devaluation (Methods) by condition, as \textbf{f}: paired and unpaired against no plasticity, each $P=6.8\times10^{-8}$, $\delta=+1.00$; paired against unpaired $P=0.36$, $\delta=+0.17$.
\textbf{h}, Valence after training (top) and devaluation (bottom) against $\eta$; dotted, 0.7; gray, zero.
\textbf{i}, The matched variant of Fig.~3: each seed's weight, valence and closest approach before (dark blue) and after training, and its devaluation, paired against unpaired $U=400$, $P=6.8\times10^{-8}$, $\delta=+1.00$.
\textbf{j}, Every seed's path after training by condition, as read; the paired seed at the median devaluation bold; green dot, the source; under each arena, the median closest approach and the seeds reaching it; bar, 10 mm.
\textbf{k}, As \textbf{j} for the matched variant.%
}

\newcommand{\mainfigure}[3]{%
  \clearpage
  \newgeometry{left=1.2cm,right=1.2cm,top=1.3cm,bottom=1.9cm}%
  \noindent{\small\textbf{Fig.~#3}}\par\vspace{0.15cm}
  \begin{center}\includegraphics[#1]{#2}\end{center}
  \clearpage
  \restoregeometry}
\newcommand{\mainfigurelandscape}[3]{%
  \clearpage
  \newgeometry{left=1.3cm,right=1.3cm,top=1.2cm,bottom=1.2cm}%
  \begin{landscape}\landscapepagenumber
  \noindent{\small\textbf{Fig.~#3}}\par\vspace{0.15cm}
  \begin{center}\includegraphics[#1]{#2}\end{center}
  \end{landscape}
  \restoregeometry}
\begin{document}
\begin{center}
{\Large\bfseries Sensorimotor behavior from an untrained brain-and-nerve-cord connectome
\par}
\vspace{0.8em}
John R. Minnick\textsuperscript{1,2,*}, Jesus Gonzalez-Ferrer\textsuperscript{2,4}, Jinghui Geng\textsuperscript{2,3},
Jason K. Eshraghian\textsuperscript{1}, Mohammed A. Mostajo-Radji\textsuperscript{2}, David Haussler\textsuperscript{2,4}
and Mircea Teodorescu\textsuperscript{1,2,4,*}\par
\vspace{0.55em}
{\small
\textsuperscript{1}Department of Electrical and Computer Engineering, University of California, Santa Cruz, CA, USA.
\textsuperscript{2}UC Santa Cruz Genomics Institute, University of California, Santa Cruz, CA, USA.
\textsuperscript{3}Department of Computer Science and Engineering, University of California, Santa Cruz, CA, USA.
\textsuperscript{4}Department of Biomolecular Engineering, University of California, Santa Cruz, CA, USA.\par
\vspace{0.3em}
*Correspondence: jrminnic@ucsc.edu, mteodore@ucsc.edu\par}

\end{center}

\section*{Abstract}
Connectomes map the neurons and synapses of an adult fly's brain and nerve cord, but how much of behavior follows from the wiring is unknown. Here we turn one such connectome into a spiking network of 172,433 neurons with a published brain model's untuned parameters, and run it alone and inside a simulated body. Two fixed rules connect them: odors and looming objects set the firing of the sensory neurons, and the firing of the leg motor neurons moves the legs. Outside the body, a pathway of known function conducts when it supplies a few percent or more of its target's input. In the body, the fly navigates around obstacles to nearly every odor source. A rewiring that keeps each neuron's connection counts and signs reaches far fewer, and the fly's leg commands replayed with the senses off reach almost none. Navigation needs the measured wiring and sensory feedback. Dopamine-gated weakening of the mushroom body's output synapses turns approach to an odor into avoidance, and the learning is associative once one class of olfactory synapse is scaled. An untrained connectome supports a behavior and its learned reversal. The pathways it cannot drive show what the wiring diagram leaves out.

\section*{Introduction}

Complete wiring diagrams exist for the nematode\cite{White1986,Cook2019} and for the brain of the fly larva\cite{Winding2023}. Electron-microscopy volumes of the adult fly, and the reconstructions made from them, cover a large part of the central brain\cite{Scheffer2020}, the whole brain\cite{Zheng2018,Dorkenwald2024,Schlegel2024} and the ventral nerve cord\cite{Phelps2021,Azevedo2024,Takemura2024}. The descending and ascending neurons that join brain and nerve cord have been matched across three of those datasets\cite{Stuerner2025}. The wiring of the adult's nervous system is known at the resolution of single synapses. Two connectomes each cover the brain and the nerve cord of a single animal. One is the Brain And Nerve Cord connectome (BANC), from a female\cite{Bates2026}, and the other is the male central nervous system connectome\cite{Berg2025}. In them a path can be followed from a sensory neuron to a motor neuron within one dataset. A connectome records which neurons are connected and by how many synapses, and each neuron's transmitter can be predicted from the images\cite{Eckstein2024}. It leaves out the dynamics of the neurons and their modulation\cite{Bargmann2013,Marder2012}.

How much of a nervous system's function follows from its wiring is an open question\cite{Bargmann2013,Lichtman2008,Zador2019,Lappalainen2024,Beiran2025}. A spiking model built from the cataloged cell types of the protocerebral bridge reproduced the ring-attractor dynamics of the fly's heading signal over a wide range of parameters\cite{Kakaria2017}. A leaky integrate-and-fire model of the fly brain, with its parameters fixed in advance and its weights taken from the connectome, predicted which neurons drive feeding and grooming, and experiments confirmed most of the predictions\cite{Shiu2024}. Simulations of the nerve cord's connectome identified a small circuit that generates the walking rhythm\cite{Pugliese2025}. Networks that take their connectivity from the optic-lobe connectome, with the remaining parameters optimized for a task, predict the responses of identified visual neurons\cite{Lappalainen2024}. But a connectome alone often leaves the dynamics of a recurrent network under-determined when the properties of the neurons are unknown\cite{Prinz2004,Beiran2025}, and in the nematode the measured propagation of signals departs from what the anatomy predicts\cite{Randi2023}. One way to measure what the wiring explains is to fit nothing, run the connectome as it is read, and record where it reproduces known biology and where it fails. Many such tests are possible in the fly, because the behavioral roles of single cell types are known. Activating most descending neurons drives a stereotyped behavior\cite{Cande2018}. Particular descending neurons turn the walking fly\cite{Yang2024,Rayshubskiy2024} or make it walk backward\cite{Bidaye2014}, looming detectors of the optic lobe drive the giant fiber that triggers the escape take-off\cite{vonReyn2017,Klapoetke2017}, and the mushroom body supports olfactory learning\cite{deBelle1994,Aso2014} through depression of its output synapses\cite{Hige2015}.

A test of a nervous system as a controller needs a body, because an animal's movements change what its senses receive and behavior emerges from the interaction of the nervous system, the body and the environment\cite{Chiel1997,LobatoRios2022}. An open project is building an integrative simulation of the whole nematode\cite{Sarma2018}. Physics-based models of the fly's body exist\cite{LobatoRios2022,WangChen2024,Vaxenburg2025}. Their controllers have been written by hand or trained by reinforcement learning\cite{WangChen2024,Vaxenburg2025}, including one that takes the brain connectome as its architecture and learns to control the body\cite{Jin2026}. What the measured wiring of the brain and nerve cord contributes when it is run untrained in such a body, inside a closed sensorimotor loop, is the question we address.

Here we build a spiking network of 172,433 neurons from the BANC connectome, with every parameter of the published brain model\cite{Shiu2024} kept at its published value, and test it in three settings. Outside the body, we drive 15 pathways with a known behavioral role and score each by criteria written before the test. A pathway conducts when it supplies about 2\% or more of its target's input, and the failures trace to what a wiring diagram lacks, an electrical synapse and ongoing activity. In the same setting the olfactory pathway as read, every synapse at its predicted sign and count, gives a dense odor code, and scaling the output of one class of antennal-lobe neuron brings it into the range measured in the fly. In the NeuroMechFly body\cite{WangChen2024}, coupled through two fixed rules, the network reaches 95\% of 32 odor sources across nine arenas, where a network rewired to keep every neuron's degrees and signs reaches 18\%. Lesion and rescue experiments place the behavioral deficit at the output of the mushroom body, and replaying the same motor commands without sensory feedback reaches 2\% of the sources. Dopamine-gated depression of the mushroom-body output synapses while the fly forages shifts the balance of that output from approach to avoidance, and the trained fly turns away from the odor. The depression is associative when Kenyon-cell activity depends on the odor.

\section*{Results}

\subsection*{The connectome as a spiking network}
We built a spiking model of the BANC connectome, an electron-microscopy reconstruction of the brain and ventral nerve cord of one adult female\cite{Bates2026}. It has 172,433 neurons, 11.75 million connections and 35.7 million synapses. The count includes 6,404 cells that carry a non-neuronal label in the connectome's table and hold 0.07\% of the output synapses (Methods). Each neuron is a leaky integrate-and-fire unit with the parameters of the published brain model\cite{Shiu2024}, and each connection is signed by its presynaptic neuron's predicted transmitter\cite{Eckstein2024} (Methods). Every parameter keeps its published value. The network runs inside the NeuroMechFly body\cite{LobatoRios2022,WangChen2024} under a fixed interface of two rules (Fig.~1a; Methods). The arena's odor and looming set the rates of the olfactory receptor neurons and of the LC and LPLC visual projection neurons. The six leg motor-neuron pools set the gait, with the turn scaled by the output of the mushroom-body output neurons (MBONs) and by the DNa02 pair. We call the balance between the MBONs that drive approach and those that drive avoidance the valence, because in the fly the MBONs encode the valence of an odor\cite{Aso2014b,Owald2015a,Owald2015b}.

The 24 regions we read from follow the connectome's labels (Fig.~1b, d; Methods). A 60 ms pulse of odor at the receptor neurons reaches the lateral horn and the descending neurons within 12 ms of its onset, the mushroom body within 18 ms and the leg motor pools within 24 ms (Supplementary Fig.~1). A looming pulse at the visual projection neurons reaches the descending neurons within 6 ms and the leg pools within 18 ms. The order in which neurons first spike follows the order in which a traversal of the wiring from the stimulated neurons reaches them (Spearman $\rho = 0.59$ over the 6,775 neurons that spike after the odor pulse and 0.58 over the 3,884 after the looming pulse; Supplementary Fig.~2). Those neurons are 4\% and 2\% of the ones the traversal reaches.

Structure and function connect the regions with different densities. Of the 576 ordered pairs of the 24 regions (each region onto every region, itself included), 358 are joined by at least one synapse and 96 by more than 10,000 (Fig.~1c, e). The largest blocks lie within a region. The descending neurons make 1.53 million synapses, 0.65 million of them onto nerve-cord intrinsic neurons. Driving each region alone at 40 Hz and reading the spiking of every other region gives a sparser map, in which 139 of the 552 region pairs respond at 0.1 Hz per neuron or more (Fig.~1f). The strongest responses include olfactory input onto the mushroom body (33 Hz per neuron) and the lateral horn (22 Hz), other central-brain input onto the mushroom body (29 Hz) and fan-shaped body input onto the noduli (30 Hz). Seven regions evoke no response elsewhere. Three are the noduli, the protocerebral bridge and the rest of the central complex, where 71 to 100\% of the synapses onto other regions come from cells signed inhibitory, which lower nothing in a silent network. The other four are the motor groups, which supply no region with as much as 0.1\% of its input. The map follows the excitatory-signed synapse counts between the regions at $\rho = 0.65$ over the 334 region pairs with any synapse, and the counts regardless of sign at 0.56 (Supplementary Fig.~3); the strongest responses with no direct synapse, the receptors' onto the mushroom body and the lateral horn, are relayed through the antennal lobe. Six regions are never reached (the optic lobe and five sensory groups, which receive almost no central input). The map is near-deterministic, with a seed-to-seed coefficient of variation of 0 at the median and above 0.01 in 83 of the 576 cells. A pulse crosses the network in the order the wiring predicts, and the map of which region drives which follows the synapse counts between them.

\subsection*{Pathway tests outside the body}
Before putting the network in a body we asked whether it reproduces pathways whose behavioral role is known from experiments (Fig.~2a, b; Methods). We drove each population outside the body and read an identified target, with the metric and threshold of every test written down before it ran. Every test has an undriven control and a twin on a rewired network that keeps each neuron's degrees and signs and shuffles its partners. A result counts only if it conducts and dies under the shuffle.

In the taste pathway, sugar drives the proboscis motor neuron MN9 and bitter suppresses the response, as in the animal\cite{Gordon2009} and in the brain model\cite{Shiu2024}. MN9 drives the muscle that protracts the rostrum\cite{McKellar2020}, and the labellar sugar receptor neurons supply 7.6\% of its 4,608 input synapses through 38 relay neurons. Driving them fires MN9 at 1.9 Hz at 200 Hz drive and not at 100 Hz or below. The rewired twin gives 0 Hz at every rate (Fig.~2c). Adding bitter receptor drive at 100 and 200 Hz lowers the response to 0.76 and 0.02 Hz, and Ir94e drive lowers it to 0 Hz at either rate. Water, the control, leaves it at 2.6 and 2.4 Hz. Bitter and Ir94e alone each leave MN9 silent, and in the animal both inhibit proboscis extension\cite{Shiu2024}. Water alone leaves MN9 silent as well, where in the thirsty animal it elicits proboscis extension and in the brain model it fires MN9\cite{Shiu2024}. Shuffling only the tastant's outgoing connections lifts the suppression (1.93 and 1.41 Hz at 100 Hz bitter and Ir94e; $P = 0.0008$ and 0.0002 over nine seeds) (Fig.~2g). We had written down before the test that we expected the suppression to fail, because the bitter path onto MN9 is mostly excitatory at one hop, and it held nonetheless.

Escape follows the synapse counts onto the giant fiber, the descending neuron whose spike triggers the fast take-off\cite{vonReyn2014}. LPLC2, which supplies 6.5\% of the giant fiber's input, drives it at 44.7 Hz, and LC4 with LPLC2 (8.7\%) at 47.7 Hz, with the rewired twin at 0 Hz in both cases\cite{vonReyn2017,Klapoetke2017,Ache2019b}. LC4 alone (2.2\%) is silent at 40 Hz drive and conducts from 80 Hz (0.5, 9.6 and 19.6 Hz above control at 80, 150 and 250 Hz; twin 0 Hz). LC6, LC15 and LPLC1 drive jumping in the animal\cite{Wu2016} but make 0 to 2 synapses onto the giant fiber and do not drive it, nor does LC11, the control (Fig.~2d, f). The giant fiber's own output onto the tergotrochanteral motor neuron is 13 chemical synapses over the two sides (8 and 5), 0.5\% of that neuron's input, and it does not conduct. In the animal the motor neuron is coupled to the giant fiber by gap junctions\cite{Namiki2018} that accompany the chemical synapses\cite{Strausfeld1983}. The motor neuron follows the giant fiber once a connection is added by hand on each side to stand for the electrical part. It does so from 104 synapse equivalents per side (7.9\% of its input with the chemical synapses included, 19 Hz), and one for one from 208 (Fig.~2f). Every pathway tested onto an identified neuron conducts at some drive if it supplies at least 2\% of that neuron's input, and the seven supplying 0.03\% or less do not (Fig.~2f).

Descending control lateralizes in the wiring alone. DNa02 turns the walking fly toward the side on which it is activated\cite{Yang2024,Rayshubskiy2024}. Driving one DNa02 at 600 Hz, one cell per side, lateralizes the leg-motor drive to the driven side (asymmetry $+0.71$ on the left, $-1.00$ on the right), and the rewired twin does not (Fig.~2e). PFL3, the central-complex steering output\cite{Westeinde2024} (8 and 17 cells per side in this connectome), lateralizes the same way ($+0.61$, $-0.39$), and PFL1, the control, never reaches the cord. DNg12 (45 cells), a command for anterior grooming\cite{Guo2022}, recruits the front-leg pools (24.6 Hz against 0.008 and 0.029 Hz in the middle and hind pools; twin 0.13 Hz). DNa01\cite{Namiki2018,Rayshubskiy2024} lateralizes only from the left in the silent network (asymmetry $+0.49$; the right cell drives no leg motor neuron). MDN, the backward-walking command\cite{Bidaye2014}, drives about eight leg motor neurons, in the front pools and the right hind pool (330 spikes in 3 s; the twin 92 spikes in one neuron). Neither setting can show whether the walking would be backward, because the leg pools carry no direction and the gait rule turns their rates into forward stepping. DNp09, whose activation drives forward walking with a turn toward its own side\cite{Bidaye2020} and, depending on the fly's state, running and freezing\cite{Zacarias2018}, gives two spikes of leg drive. LC16, whose retreat runs through MDN\cite{Sen2017}, gives no leg drive at 40, 600 or 1,500 Hz, and LC10a\cite{Ribeiro2018} conducts but survives the shuffle.

With a background of independent 2 to 10 Hz input to every neuron, LPLC2 still drives the giant fiber (24.5, 14.2 and 5.2 Hz above control at 2, 5 and 10 Hz), and DNg12 still drives the front legs (9.8, 8.5 and 6.1 Hz, 94 to 100\% of it in the front pools). The sugar response falls to 0.35 Hz at 2 Hz and no longer exceeds its threshold (Fig.~2i). At 20 Hz only DNg12 still does. On that background the steering neurons act by suppression (Fig.~2i). At 2 Hz, DNa01 on the left lowers the drive to its own side by 1.91 Hz and raises the other side by 0.82 Hz (twin 0.06 Hz). DNa02 on the left lowers its own side by 1.35 Hz and on the right by 1.29 Hz (twins $+0.02$ and $-0.15$ Hz); every seed falls below its control, but the left misses the registered threshold of three control standard deviations, 1.35 Hz, by 0.004 Hz and the right by 0.055 Hz. Bilateral DNa02 lowers the total by 2.27 Hz (twin $+0.54$). The animal shows the same sign, with DNa02 shortening the strides of the legs on its own side and its bilateral activation slowing walking\cite{Yang2024}. DNp09 exceeds its threshold at 2 Hz, with suppression (2.21 Hz below control; twin $-0.20$). The suppression is the opposite of the forward walking registered as its sign\cite{Bidaye2020}, and it is compatible with the freezing that its activation also induces\cite{Zacarias2018}. MDN also conducts with suppression (1.70 Hz below; twin $+0.69$). Of the 15 pathways, seven conduct, die under the shuffle and meet the criterion registered for the animal's behavior: sugar, bitter, LPLC2, LC4 with LPLC2, DNa02, PFL3 and DNg12 (Fig.~2h). The criterion for DNa02 and PFL3 is a higher leg-motor drive on the driven side, whether the left or the right is driven. In the body a higher drive on one side turns the fly away from that side, where the animal turns toward it (Fig.~5j). The fall in drive that would turn the fly toward the driven side appears for DNa02 only on a background, just under the registered threshold. The giant fiber's output passes only with the gap junction added, DNa01 only on a background and driven from the left, and DNp09 only on a background and with the sign opposite to the one registered. MDN conducts and dies under the shuffle, but the model cannot read its direction. The four that fail the test are leg sugar, which supplies 0.02\% of MN9's input, LC4 alone at 40 Hz, LC16 and LC10a. A pathway conducts when it supplies about 2\% or more of its target's input. Each conditional pass marks something the model lacks, whether an electrical synapse, ongoing activity or a readout of walking direction.

\subsection*{The odor code outside the body}
The 2,978 receptor neurons reach 616 of the 699 projection neurons through 203,937 synapses and the antennal lobe's 429 local neurons through 172,141 (Fig.~3a). The projection neurons return 217,657 synapses onto the local neurons and send 168,315 onto the 4,504 Kenyon cells. The Kenyon cells' 28,415 synapses onto the 104 MBONs include the ones the learning of Fig.~6 changes. The transmitter prediction makes the local neurons 161 glutamatergic, 146 GABAergic, 105 cholinergic, 14 serotonergic and 3 octopaminergic (Fig.~3b). The cholinergic class makes the most synapses, with a median of 3,029 output synapses per cell against 456 for the GABAergic and 304 for the glutamatergic cells. In the fly the local neurons are diverse in transmitter\cite{Chou2010}. Many are GABAergic\cite{Wilson2005}, a third are glutamatergic and inhibit\cite{Liu2013}, and excitatory ones exist\cite{Shang2007,Huang2010}. The model has the APL neuron, the feedback that sparsens the Kenyon-cell code in the animal\cite{Lin2014}. It consists of two cells, one of them without a label in the neuron table, with 22,926 synapses onto the Kenyon cells. Both cells excite the Kenyon cells in the model because they are predicted cholinergic, although the APL is GABAergic in the fly\cite{Liu2009a} and GABA is the transmitter the connectome's own annotation lists as verified for it (Methods). Silencing the labeled APL lowers the share of Kenyon cells active from 68\% to 61\%.

The network as read gives a dense code. We presented two odors of five disjoint glomeruli each (Fig.~3c). Either odor fires 68\% of the Kenyon cells, and the two codes' 5\% most active cells overlap by 0.90. The projection-neuron layer is already 74\% active, with an overlap of 0.86, and the MBONs fire about 2,400 spikes per presentation whatever the odor. In the fly, 5 to 10\% of Kenyon cells respond to an odor\cite{Turner2008,Honegger2011,Lin2014}. The cholinergic local neurons' output onto the projection neurons sets the density. Scaling it by a gain $g$ from 1 to 0 lowers the share of Kenyon cells active through 43\%, 15\%, 6.1\%, 3.4\%, 2.7\% and 2.3\% to 1.8\%, and the share of projection neurons active from 74\% to 10\% (Fig.~3d). The overlap falls from 0.90 to 0.24 at $g = 0.1$ before rising again in the near-silent code. We set the fly's window at 3 to 15\% before the sweep ran, wider than the measured 5 to 10\% (Methods). At $g = 0.15$ and 0.1 the code lies in that window with an overlap under 0.5, and $g = 0.15$ is the matched variant used from here on. Signing every local neuron inhibitory instead gives 1.4\% of the Kenyon cells active, below the window (Supplementary Table~4). Over eight odors the share of Kenyon cells active is 68 to 70\% as read and 4.9 to 17.2\% in the matched variant, median 6.8\% (Fig.~3e). The matched MBON output is 52 to 287 spikes per presentation, against 2,352 to 2,483 as read (Fig.~3f). Part of the matched output does not pass through the Kenyon cells. For the first odor pair, the 52 spikes to odor A (the median over six seeds) all come from one MBON05 cell. Depression of the Kenyon-cell synapses cannot remove that output, because the unlabeled APL cell drives that cell through 62 synapses while the Kenyon cells supply 154 of its 468 input synapses.

The dense code comes from the loop between the projection and local neurons. The as-read network fires 67\% of its Kenyon cells after the warm-up with no receptor input at all, 68\% at 5 Hz and 77\% at 120 Hz. It fires none when started from silence. Receptor input at 5 Hz switches it on, and it then stays on (67\% once the input is removed). The matched variant follows its input, with 4.2\%, 8.2\%, 43\% and 68\% of its Kenyon cells firing at 0, 5, 40 and 120 Hz and 4.2\% again after (Fig.~3g). Its code stays sparse for odors of up to ten glomeruli (6.0\%, 16\% and 35\% at 5, 10 and 20; as read 68\%, 70\% and 72\%). Halving or quartering the odor drive leaves the density unchanged (matched 5.3\% and 5.1\%) and raises the matched overlap (0.55, 0.73) while lowering its output (Fig.~3k). We silenced the cholinergic types one at a time (Fig.~3i, j). Silencing lLN1\_bc (34 cells, 105,956 output synapses) lowers the share of Kenyon cells active from 68\% to 50\%, and the other ten types leave 66 to 68\% active. The fall scales with the type's output (Spearman $\rho = 0.78$, $P = 0.0045$, 11 types). Types silenced together do more than the sum of their single falls. The three largest together leave 25\% of the Kenyon cells active where the sum of their single falls predicts 46\%. All eleven leave 3\% against a predicted 40\%, and all 105 cells 2\%.

Training on odor A in the matched variant depresses A's MBON output and, in part, the untrained odor's. A's output falls by 0.31 to 0.36 at training strength 1 for a drive-matched pair in which the untrained odor, B, shares $k$ of A's five glomeruli (Fig.~3h). B's depression as a fraction of A's is 0.13, 0.10, 0.49, 0.70 and 0.96 at $k = 0$ to 4 at strength 0.5, and 0.28, 0.19, 0.86, 0.96 and 1.02 at strength 1, while the Kenyon-cell overlap rises from 0.26 to 0.72. As read, both odors fall by 0.18 to 0.19 whatever they share. The depression generalizes with the overlap of the two codes, as the depression\cite{Hige2015} and the learned avoidance\cite{Campbell2013} do in the fly. By three shared glomeruli the second odor takes on 0.70 to 0.96 of the trained odor's depression, and an odor that shares no glomerulus takes on 0.13 to 0.28. The network as read gives a dense, self-sustaining code, and a Kenyon-cell code in the fly's range needs one class of antennal-lobe synapse scaled to 0.15 of its face value.

\subsection*{Navigation in the body}
Twenty flies, each a seed with its own input and initial conditions, walked nine arenas holding 32 odor sources behind walls of increasing clutter, under three wirings: intact, rewired, and with the mushroom body lesioned (Fig.~4a to d). The intact fly reached 30.25 of the 32 sources per fly (95\%). It reached every source of its arena in 161 of 180 runs and at least one in every run (Fig.~4f, g). The rewired fly reached 5.80 (18\%; no source in 92 of its 180 runs, every source in 2) and the lesioned fly 3.45 (no source in 117 runs, every source in none). A correlated random walk at the fly's own speed reached 2.20. No rewired fly reached as many sources as any intact fly (at most 11 against at least 27 of 32). The rewired fly reaches more than the lesioned one (1 to 11 against 1 to 6 per fly; Mann--Whitney $P = 0.0099$, Cliff's $\delta = +0.48$, 95\% interval $+0.14$ to $+0.76$). Its MBONs still fire, 7.5 spikes per step against the intact fly's 154, and 75 of the 104 MBONs drive approach. The balance reads a weak positive valence (0.11 against 1.41). For scale, a reference controller that is given the position of every source and walks straight to the nearest at the same speed reaches 29.00 if walls stop it and 32.00 if it slides along them (Supplementary Table~1). The intact fly's misses are mostly in the two most cluttered arenas (17 of the 19 runs that miss a source). In the four-wall slalom 9 of 20 runs reach all five sources and 5 stop at the second; in the dense slalom 14 of 20 reach all four. Along a corridor the intact fly reached the first source in every run, and the rewired and lesioned flies in 35\% and 28\% of runs (Fig.~4h). The intact fly reached a second source in 179 of its 180 runs, and the rewired and lesioned flies in 25 of the 88 and 6 of the 63 runs that reached any source. Both manipulated flies keep walking (Fig.~4i to k). The median speed is 8.9, 12.8 and 11.8 mm/s for the intact, rewired and lesioned fly, with 97.2\%, 99.6\% and 99.2\% of time moving, and the straightness 0.61, 0.31 and 0.61. The rewired and lesioned flies leave the arena frame in 64\% and 65\% of samples. The olfactory input is the same in all three (5.92 spikes per receptor neuron per step, 99.7\% of them active).

The firing rates are the network's own, measured during the runs (Fig.~4e, l, m). Every neuron's spikes were recorded in one run per condition (arena i, seed 0). In that run the intact mushroom body fires 1.93 spikes per neuron per 50 ms step with 77\% of its neurons active, and the lateral horn 1.30. The lesion removes the mushroom body and leaves the lateral horn at 1.30. The rewiring leaves the mushroom body silent (0.006 spikes per neuron, 2\% active) and the lateral horn at 0.018 under the same receptor drive. The descending and leg-motor rates stay near intact in the lesioned fly and fall in the rewired one (front-leg motor 0.13, 0.03 and 0.13 spikes per neuron per step). The MBON output, logged in every run, shows the same across all 180 runs per condition, with 154 spikes per step in every intact run (153.7 to 154.9), 7.5 in the rewired fly (3.0 to 18.0) and none in the lesioned. With the same rules and the same input, the rewired fly, every degree and sign kept, reaches 18\% of the sources where the intact fly reaches 95\%. The navigation comes from the measured wiring.

\subsection*{Perturbations in the body}
In a forage arena with three sources behind a wall (Fig.~5a to d), the intact fly reaches all three in all 20 seeds (20 flies with their own input and initial conditions, as in the battery). With the mushroom body lesioned it reaches none in 20 of 20. Its leg motor rate (152 against 154 Hz), wall avoidance (1.9 wall-contact steps) and path length (220 against 193 mm) are kept, and its looming-driven DNa02 rate falls (0.81 against 1.99 spikes per step), a loss of approach with walking spared. The rewired fly reaches 0.10 of 3 (18 of 20 none) and loses more, with its DNa02 drive at 0.54 and its leg motor rate at 98 Hz (64\% of intact). Pinning the valence scalar at 1.44, close to the intact fly's 1.41, restores approach in both, to 3.00 and 2.55 of 3 (20 and 17 of 20 all three). It does not restore the rewired fly's walking (103 Hz) or the lesioned fly's DNa02 rate (0.77). The behavioral deficit of both manipulations lies at the mushroom-body output. The rescue has a threshold below 0.7, a valence at which the lesioned fly is rescued as fully as at 1.44 (3.00 of 3). The intact fly pinned at 0 reaches nothing while walking normally (leg motor rate 99\% of intact). A random lesion of 4,608 neurons, the size of the mushroom-body set, leaves approach mostly intact at 2.55 of 3, with 3 of 20 seeds reaching nothing against the lesion's 20 of 20 (Fisher's exact $P = 2.6 \times 10^{-8}$). It costs walking (leg motor rate 85\% of intact). Cutting the visual loop, with the looming signal set to zero, abolishes wall avoidance (233 wall-contact steps against none in the intact fly). In this walled arena that also abolishes approach (0.30 of 3). The leg motor rate is kept (99\% of intact), but the fly stays at the wall and its path falls to 53 mm from 193. Supplementary Table~2 gives every measure for every condition.

Cutting the whole sensory loop abolishes navigation (Fig.~5e, g). Each closed-loop run's own motor commands were replayed to the body from a start rotated by 30$^{\circ}$, with the network's output ignored. The replays reach 0.07 sources per run against 3.29 for the closed loop from the same rotated start (2.0\% against 94.0\% of an arena's sources, averaged over the runs). Of the 180 replays, 169 reach nothing and none reaches every source, while 154 of 180 closed-loop runs reach every source. All 20 flies, all nine arenas and all 180 pairs fall the same way ($W = 0$, $P = 1.9 \times 10^{-6}$ over flies). The rotation alone does not account for this, because the rotated closed loop reaches 3.29 sources per run against 3.36 for the unrotated record ($P = 0.13$).

Driving one descending neuron in the body reproduces the sign measured in the animal where the wiring did (Fig.~5f; Supplementary Table~3). Every turn rule of the interface is switched off in these runs, and the body follows the leg-motor pools alone. DNa01 turns the fly $+116^{\circ}$ from the left and $-150^{\circ}$ from the right against a no-drive baseline of $-7^{\circ}$ and $+0.5$ mm/s over the 2 s of drive, a mirror the rewired twins lack ($+55^{\circ}$ and $+36^{\circ}$). The threat-response cluster slows it by 5.1 mm/s where, rewired, it speeds it up by 6.3. DNa02 turns the fly right when driven bilaterally ($-38^{\circ}$ against $-7^{\circ}$ for no drive) and, within the no-drive spread, from either side alone ($-16^{\circ}$ and $-25^{\circ}$). Its bilateral drive speeds the fly up (0.8 mm/s) where the animal slows\cite{Yang2024}. DNp09 turns left ($+54^{\circ}$) and slows (2.2 mm/s) where the animal walks forward\cite{Bidaye2020}, and DNp01, the giant fiber, does nothing. DNg12 speeds the fly up by 1.8 mm/s, as does its twin (2.2), because the gait controller reads its front-leg drive as amplitude. The sign depends on a baseline (Fig.~5j). Every drive speeds the silent fly up with the receptor drive at 0 Hz, and the steering neurons turn it away from the driven side (DNa01 left $-219^{\circ}$, DNa02 left $-269^{\circ}$ and right $+219^{\circ}$), the excitation-only regime of the silent wiring (Fig.~2). DNa01 turns the fly toward the driven side from both sides with 40 or 120 Hz of receptor drive ($+86^{\circ}$ and $-167^{\circ}$ at 40 Hz), and DNa02 does so from the right ($-47^{\circ}$ against $+21^{\circ}$ for no drive at 40 Hz). The threat cluster arrests the fly (5.2 and 5.1 mm/s) where at 0 Hz it accelerated it (3.1). The speed increase reported for DNp09 appears only in the silent body (1.7 mm/s) and reverses on a baseline ($-2.1$ and $-2.2$ mm/s at 40 and 120 Hz).

Navigation degrades gradually when neurons are removed at random, and abruptly when the neurons with the most synapses are removed first (Fig.~5h, i). Random removal of 5\%, 10\% and 20\% of the candidate neurons leaves 2.53, 2.20 and 2.50 of 3 sources, 30\% leaves 1.30, 50\% 0.30 and 70\% 0.18, with the leg motor rate falling from 123 to 4 Hz. Removing the 1\% of neurons with the most synapses leaves no source reached, with the leg motor rate at 6 Hz, and so does removing 2\%. A run that ends in a physics fault, the partly driven body falling, is left out of its mean. Eleven of the 20 runs fault at 1\%, 2, 2 and 3 at 2\%, 4\% and 6\%, and none from 8\% on, where the legs have stopped. Random removal loses 1 and 3 of its 20 runs to faults at 5\% and 70\% (22 of 280 runs in the two arms). Removing the top 1\% within the central brain alone (312 neurons) leaves 0.78 of 3 in 18 completed runs with the leg motor rate at 15\% of intact, and from 2\% navigation is gone with the rate at zero. Ranking by traffic reaches the premotor drive before any navigation-specific core, and the test cannot separate navigation from walking. Both manipulations lose approach at the mushroom-body output. Navigation also needs the sensory loop, without which the same leg commands reach on average 2\% of an arena's sources against 94\% in the closed loop.

\subsection*{Learning in the loop}
We let the Kenyon-cell-to-MBON synapses of the aversive compartments depress under a reinforcement signal while the fly forages, with nothing else changed (Fig.~6a; Methods). The animal learns aversive olfactory associations by depression of the same synapses\cite{Hige2015}. Dopaminergic neurons deliver the reinforcement\cite{Schwaerzel2003,ClaridgeChang2009}, and the learned behavior follows from a shift in the balance between the MBONs that drive approach and those that drive avoidance\cite{Aso2014b,Owald2015b}. Six training episodes with the odor paired with the reinforcement take the mean weight of the 8,377 plastic connections from 0.513 to 0.138 mV. The approach-driving MBONs fall from 103 to 37 spikes per step while the avoidance-driving MBONs stay at 50 (49 after training), and the valence goes from 1.41 to $-0.33$ (Fig.~6b, d, e). One training episode is enough to take it there. Before training every fly reaches the source. After training none of 20 does (Fig.~6f, g, j). The trained fly turns away from the odor, and its closest approach to the source is a median 19.6 mm against 0.15 mm without plasticity ($P = 6.8 \times 10^{-8}$, Cliff's $\delta = +1.00$ against no plasticity, on closest approach and on devaluation). The learning rate is a dose with a threshold. At $\eta = 10^{-5}$ the valence after training is 0.44 and all 20 flies still reach the source; at $3 \times 10^{-5}$ it is $-0.21$ and none does (Fig.~6b, h). From $10^{-4}$ to 0.08 it stays between $-0.31$ and $-0.33$ while the weights go on falling, from 0.233 to 0.134 mV. The valence levels off because the depression is confined to its targets. In the recorded seed the 41 approach-driving MBONs of the aversive compartments fall from 69 to 6 spikes per step. The 34 approach-driving MBONs of the other compartments (35 to 31) and the avoidance-driving MBONs (50 to 49) keep their output (Fig.~6d).

In the network as read the depression is not associative. Reinforcement with the odor withheld depresses the synapses almost as much (weight 0.183 mV, valence $-0.29$). Those flies turn away as well, with none of 20 reaching the source ($P = 0.49$ against the paired flies on closest approach, $\delta = +0.13$, 95\% interval $-0.25$ to $+0.49$). The as-read Kenyon cells fire without the odor (67\% of them at 0 Hz of receptor input and 68\% at the 5 Hz the arena supplies when the odor is withheld, Fig.~3g; 69\% in the body, Fig.~6c), and the eligibility trace is present with or without it. In the matched variant few Kenyon cells fire without the odor (4\% at 0 Hz and 8\% at 5 Hz, Fig.~3g; 12\% in the body, Fig.~6c). There the same protocol is associative at the synapse and in the behavior. Pairing takes the weight to 0.195 mV and the valence to $-0.53$, and none of 20 trained flies reaches the source (median closest approach 21.1 mm). Reinforcement alone leaves the weight at 0.508 mV and the valence at 1.38, and all 20 flies reach the source, as they do without plasticity (paired against reinforcement alone, $P = 6.8 \times 10^{-8}$, $\delta = +1.00$) (Fig.~6i, k). The registered tests, which are on the behavior, read associative and expressed (Methods). Both codes are dense while the odor is present (77\% as read, 68\% matched, every receptor neuron driven). The depression is associative when the code depends on the odor, whether or not the code is sparse.

The avoidance needs both the balance readout and the confined reinforcement; two further arms change one of the two at a time (Supplementary Fig.~4; Supplementary Table~7). A sum over every MBON cannot change sign. With the valence read as the summed spike count of every MBON, and every connection plastic, the same depression lowers it from 1.44 to 0.24. That floor is held by MBONs the Kenyon cells do not drive. Of 20 trained flies, 19 then still reach the source (Supplementary Discussion; Supplementary Table~6 sets the two readouts side by side). With the balance and the reinforcement in every compartment the trained fly walks past the source, steering neither toward it nor away (valence after pairing 0.03, $-0.02$ in the matched variant; 2 and 1 of 20 flies reach the source). With the summed count and the confined reinforcement all 20 flies still reach it. In the loop, dopamine-gated depression of one synapse class turns approach into avoidance. It is associative in the matched variant, where few Kenyon cells fire without the odor, and not in the network as read, where most do.

\section*{Discussion}

We built a spiking network from a complete brain-and-nerve-cord connectome with every parameter at its published value, tested it against pathways of known behavioral role, placed it in a body and perturbed it there. A pathway of known behavioral role conducts in the untrained network when it supplies about 2\% or more of its target's input; those that fail supply a hundredth of that or depend on something the wiring diagram does not record. In the body the intact wiring reaches 30 of 32 odor sources. Plasticity at one class of synapse, run inside the loop, turns the fly's approach to an odor into avoidance.

Each result of the network as read is a property of the wiring, because none of its parameters was fitted. Other models optimize a connectome's unknown neuron and synapse parameters against a task\cite{Eschbach2020,Lappalainen2024}, train a deep network against the behavioral effects of silencing identified neurons\cite{Cowley2024}, or train a controller with the connectome as its architecture by reinforcement learning\cite{Jin2026}. These approaches reach an accuracy that an untrained model does not attempt. A trained interface can make a network drive a body whatever the wiring. When a worm connectome was read into a fly body through a decoder trained by reinforcement learning to imitate real flies, the body walked like a fly\cite{Brunton2026}. Our interface has no trained part, only a fixed input rule and a fixed readout rule. Under the same rules the rewired network, with every degree and sign kept, finds few of the sources. Once parameters are fitted, how much the wiring alone explains can no longer be asked.

The giant fiber fails to drive its motor neuron because in the animal the two are coupled by gap junctions\cite{Namiki2018} that accompany the chemical synapses\cite{Strausfeld1983}, and the connectome holds only the chemical part. A connection added by hand on each side restores the pathway. The steering neurons need ongoing activity to act on. In the walking fly DNa02 shortens the strides of the legs on its own side\cite{Yang2024}, and the steering signals of the two hemispheres change in opposite directions\cite{Rayshubskiy2024}. A silent network cannot lower a rate that is already zero. The driven side's leg drive is suppressed in the network with 2 Hz of background, and DNa01 turns the fly toward the driven side from either side once the body has receptor drive. Command-like descending neurons also act in the animal by recruiting further descending neurons\cite{Braun2024}, which may be why one cell driven alone gives part of a behavior. The share of a target's input that a pathway supplies says where anatomy is likely to be enough. The 2\% boundary is a result of this model at its published synaptic weight and threshold\cite{Shiu2024}.

Bates et al.\ describe the architecture of this connectome, from a linear measure of influence, as local loops from sensory to effector neurons, linked by ascending and descending neurons, with the regions for learning and navigation supervising them\cite{Bates2026}. The spiking map of Fig.~1f is the dynamical counterpart of that analysis. It is sparser than the anatomy, 139 responding region pairs against 334 connected ones, because a spiking network passes on only the input that reaches threshold. The embodied results agree with the supervisory reading. Removing the mushroom body removes approach and leaves walking and wall avoidance intact, where a random lesion of the same size leaves approach mostly intact.

The body is coupled to the network by two hand-written rules with fixed constants, and they set the scope of the embodied claims. The turn toward or away from an odor is computed from the odor's left-right difference and scaled by the valence, the balance between the MBONs that drive approach and those that drive avoidance. The battery, the lesion, the rewiring and the rescue test what the connectome supplies to that rule, the valence. That quantity reaches the MBONs through four layers of measured connectivity, and it falls to less than a tenth of its value when that connectivity is shuffled (0.11 against 1.41). Which MBONs the valence reads is part of the rule. Under a sum over every MBON the lesion, rescue and replay results are the same and the learning has no effect on the behavior (Supplementary Discussion; Supplementary Table~6). The rescue places the deficit of both manipulations at the mushroom-body output, without showing that the connectome computes the turn. The receptor drive has a fixed total rate, and the network is never asked to encode concentration. The gait is the body model's, with amplitudes set by the leg motor pools. Three results are independent of the odor rule: the pathway tests outside the body, the descending-neuron drives in the body, run with every turn rule switched off, and the replay experiment. Replay holds the motor commands fixed and removes only the feedback. Navigation in this body depends on the sensorimotor loop, whatever the origin of the commands. Locomotor control in animals likewise combines feedforward commands with sensory feedback\cite{Dickinson2000}.

The antennal lobe is where reading the connectome at face value departs furthest from the animal. In the fly the excitation of projection neurons by the cholinergic local neurons\cite{Shang2007} passes through electrical synapses\cite{Yaksi2010}, although paired recordings also find a cholinergic component\cite{Huang2010}. Between glomeruli, inhibition dominates over excitation\cite{Olsen2008}. The model has no electrical synapses and reads each chemical synapse of these cells as full-strength excitation. This closes a loop between projection and local neurons that sustains itself once started, and two thirds of the Kenyon cells then fire to every odor. Scaling that one class of synapse to 0.15 of its face value brings the Kenyon-cell code into the fly's range. The APL neuron, whose feedback keeps the code sparse in the fly\cite{Lin2014}, is GABAergic in the animal\cite{Liu2009a} and cholinergic in the transmitter prediction. We kept the prediction for every neuron rather than correcting cells one at a time, because the model's claim is about the connectome as read. Signing the two APL cells GABAergic, or applying every verified sign in the neuron table, lowers the share of Kenyon cells active less than scaling the loop does and leaves it above the fly's range (Methods). The model's prediction for learning is that depression of the Kenyon-cell output synapses is associative when Kenyon-cell activity depends on the odor. Where the code does depend on the odor, generalization between odors follows the overlap of their codes, as it does in the fly\cite{Campbell2013,Hige2015}. In the loop the depression of the input to approach-driving MBONs tips the balance of MBON output toward avoidance, and the trained fly turns away from the odor, as aversively trained flies do\cite{Aso2014b,Owald2015b}.

The model's assumptions are few: point neurons with one sign each, chemical synapses only, no neuromodulation, no spontaneous activity, one animal. Neuromodulators reconfigure what a fixed wiring diagram does\cite{Marder2012}, and the strength of a connection depends on the density and the location of its synapses as well as on their number\cite{Liu2022a}. The synapse detection of this connectome finds about a fifth as many output synapses per Kenyon cell as an earlier whole-brain volume\cite{Bates2026}, and the valence and the learning reported here were obtained with that count. Each omission can be added back on its own and scored against the same measurements, because the battery and the pathway tests are fixed. Candidates are a background matched to the activity across the fly's brain that accompanies walking\cite{Brezovec2024}, electrical synapses as they are mapped, and synaptic weights that differ by transmitter. Another is a reinforcement signal produced by the network's own dopaminergic neurons, whose inputs include feedback from the MBONs\cite{LiFeng2020}, in place of an imposed one. A causal model of the fly brain has been proposed that uses the connectome as a prior for perturbation data\cite{Pospisil2024}; the input-share rule and the functional map indicate where that prior is already informative.

Three predictions follow for experiments in the animal. A pathway should drive its target when it supplies a few percent of the target's input synapses and fail at a hundredth of that. Steering by DNa01 and DNa02 should depend on ongoing activity in the leg motor network, because it acts by lowering that activity on one side. A fly whose Kenyon cells are active without an odor should be devalued by reinforcement that is not paired with the odor.

\section*{Methods}

\subsection*{The connectome and the network model}
From the BANC connectome of the adult female \textit{Drosophila} brain and ventral nerve cord\cite{Bates2026} we used the neuron table, the connection list (neuron pairs with their synapse counts) and the per-neuron transmitter prediction. Every neuron that appears in the connection list is simulated, 172,433 neurons joined by 11,752,828 connections that carry 35,733,096 synapses. Of these, 6,404 carry a non-neuronal label in the table (6,281 glia, 83 trachea, 40 marked as not a neuron); they were left in as units like the rest, hold 0.07\% of the output synapses, and fire 0.003\% of the spikes of an embodied run. The connection list names each neuron by its identifier in the release it was built from. The neuron table carries that identifier beside a newer one from the latest release, which differs for 12,806 neurons whose segments changed between releases; the table is joined to the list on the build identifier. No neuron, connection or synapse was added, removed or reweighted except where a manipulation below says so. The list includes 156,311 connections from a neuron onto itself (1,422,866 synapses, 4.0\% of all). The connectome's authors expect most connections of this kind to be errors of synapse assignment and recommend filtering them out\cite{Bates2026}. They were left in here; they do not act on a neuron that has a refractory period, because their input arrives 1.8 ms after the neuron's own spike, inside the 2.2 ms period, when synaptic input is discarded. The authors also note that their synapse detection can miss some classes of connection. A Kenyon cell has 36.6 outgoing synaptic links on average, against about 200 in an earlier whole-brain volume, and no correction was made for this\cite{Bates2026}.

Each neuron is a leaky integrate-and-fire unit with the equations and parameters of the brain model of Shiu et al.\cite{Shiu2024}, who took them from earlier fly models and recordings except the synaptic weight, their one free parameter, which they set so that sugar input drives the proboscis motor neuron and showed robust to a 30\% change; we retuned none of them. The membrane potential $v$ and the synaptic variable $g$ of each neuron obey
\begin{equation}
\tau_m\,\frac{\mathrm{d}v}{\mathrm{d}t} = v_0 - v + g, \qquad \tau_s\,\frac{\mathrm{d}g}{\mathrm{d}t} = -g,
\end{equation}
with $v_0 = -52$ mV, $\tau_m = 20$ ms and $\tau_s = 5$ ms. A spike is emitted when $v$ exceeds $-45$ mV; $v$ is then reset to $-52$ mV, $g$ to zero, and the neuron is refractory for 2.2 ms. A presynaptic spike adds to the postsynaptic $g$, 1.8 ms later,
\begin{equation}
\Delta g = s\,n \times 0.275\ \mathrm{mV},
\end{equation}
where $n$ is the number of synapses of the connection (synapse counts approximate synaptic contact area\cite{Barnes2022}) and $s$ is the sign of the presynaptic neuron: $+1$ for acetylcholine, dopamine, serotonin, octopamine and tyramine, $-1$ for GABA, glutamate and histamine. Glutamate is signed as inhibitory, as in the brain model\cite{Shiu2024} and as measured in the fly's antennal lobe\cite{Liu2013}. The prediction is the one distributed with the connectome; predictions of this kind come from a classifier trained on electron-microscopy images of synapses\cite{Eckstein2024}. The 7,213 simulated neurons without a prediction have their 10,000 connections (29,865 synapses) signed $+1$. The network was simulated in Brian 2 (version 2.7.1)\cite{Stimberg2019} at its default time step of 0.1 ms.

Input is delivered as independent Poisson spike trains, one per driven neuron. Each event depolarizes its target by 68.75 mV ($250 \times 0.275$ mV), and the neuron fires at every event that does not fall in its refractory period. The olfactory receptor neurons and the LC and LPLC neurons, through which the body's senses enter, have no refractory period and follow their drive spike for spike, as do the other sensory receptor neurons in the runs of Fig.~1 f and Fig.~2; their self-connections (66 synapses at most) stay below the roughly 160 synapses that fire a neuron on their own. Every other driven neuron keeps its refractory period, and a drive of rate $r$ gives $r/(1 + 0.0022\,r)$ spikes per second, 259 at 600 Hz.

Populations are selected by the connectome's labels. Of the 2,978 olfactory receptor neurons, the 2,968 with a labeled nerve are driven, split by the side of the nerve they enter; the 4,192 visual projection neurons of the LC and LPLC types are split by the position of their root point relative to the midline; the 104 mushroom-body output neurons (MBONs), the DNa02 pair and the descending neurons are selected by cell type; the 391 leg motor neurons form six pools by leg and side (63 to 70 neurons each). The mushroom-body set is the 4,504 Kenyon cells and the 104 MBONs. A lesion sets to zero every synapse onto or from the set. The rewired null permutes the postsynaptic ends of all connections with a seeded random generator. Every neuron keeps its in-degree, its out-degree and the sign and synapse count of each of its outgoing connections while the identity of every partner is randomized, and each trial seed gives a different permutation. A pathway-scoped shuffle (Fig.~2 g) permutes the postsynaptic ends of one population's outgoing connections among themselves, leaving every other connection intact.

The 24 regions of Fig.~1 come from 13 groups defined by the connectome's super-class, cell-class and region labels (Supplementary Table~8 gives each group's defining labels and neuron count; the antennal-lobe projection and local neurons and the 8,409 neurons matching no rule fall in the other central-brain group), then split the central complex by cell-type prefix into its neuropils, the sensory group by cell class and the leg motor neurons by segment.

\subsection*{Stimulation outside the body}
For the functional influence map (Fig.~1 f), every neuron of one region received 40 Hz input for 3 s with no other input, and spikes were counted in every neuron in 20 ms steps. The response of a region is its mean rate per neuron over the episode; each of the 24 regions was driven in turn, three seeds each. For display, each row was transformed as $\log(1 + \text{rate})$ and divided by its maximum. Rows and columns of Fig.~1 e and f follow average-linkage clustering of the symmetrized log synapse counts.

For the propagation snapshots (Supplementary Fig.~1) the network was run for 60 ms with no input, and spikes were then counted in 6 ms bins for 144 ms; from 18 to 78 ms of that record the receptor neurons of both sides received 400 Hz (odor) or every LC and LPLC neuron of both sides received 300 Hz (looming). The rows of the figure are the bins starting 0, 6, 12, 24 and 60 ms after the onset, and a region's color is its rate in the bin, spikes per neuron per 6 ms, on a logarithmic scale from 0.001 to the input layer's peak, over a gray outline of every region. A region is reached in the first bin in which its spikes per neuron are at least one tenth of their mean over the pulse, and the time given is the end of that bin. For Supplementary Fig.~2 the wiring was traversed from each pulse's stimulated group with the information-flow model of Schlegel et al.\cite{Schlegel2021}, in the form described for the FlyWire connectome\cite{Dorkenwald2024}. In each round every untraversed neuron is traversed with probability $\min(1, r/0.3)$, $r$ being the share of its input synapses that come from traversed neurons. A neuron's traversal distance is the round in which it was traversed, averaged over ten seeded traversals. A neuron's first spike is the end of the first 6 ms bin after the onset in which it spikes; the stimulated neurons are left out. For Supplementary Fig.~3 the synapse counts between the 24 regions were split by the sign the model gives the presynaptic neuron, and the influence of Fig.~1f was compared with them per region pair by Spearman's $\rho$ over the pairs with any synapse. Fig.~1 a, b and d draw the connectome's neuron skeletons.

\subsection*{Validation against pathways of known behavioral role}
Each test of Fig.~2 drives a population with a known behavioral role and reads an identified target neuron or motor pool; the rows are those of Fig.~2 a, b and h, cited in the Results. LC and LPLC populations were driven at 40 Hz per neuron (LC4 also at 80, 150 and 250 Hz; LC16, LC10a and LC10d also at 600 and 1,500 Hz); descending neurons, singly or as a type, and the PFL3 and PFL1 populations at 600 Hz, which their refractory period turns into 259 spikes per second; and the taste receptor neurons over 10 to 200 Hz, with the two-tastant grids scored at 200 Hz sugar. LC10d was run as a replicate of LC10a and is not among the 15 scored pathways. Each run lasted 3 s with spikes counted in 20 ms steps; the readout is the target's mean rate over the run, and for the descending rows the rate of each leg-motor pool by side. Every driven condition was run with an undriven control at the same settings and with a twin on the rewired network whose permutation seed equals the run seed; each had nine seeds. The background ladder (Fig.~2 i) adds Poisson input to every neuron at 0, 2, 5, 10 or 20 Hz, and effects are measured against the control at the same background.

A pathway conducts when the driven target's rate exceeds its control's by at least 0.10 Hz and by the row's threshold, which is half the control's own rate, or three seed-to-seed standard deviations of the control for the sided rows. It is wiring-specific when the twin's effect is under half the intact effect. A sided row must also have the higher leg-motor drive on the driven side, whether the left or the right is driven, the measure registered for the animal's turn toward that side. A side is not read when its total leg-motor drive is under 0.10 Hz. In the model's body the higher drive turns the fly away from the driven side, because the gait rule gives the legs with the higher drive the larger amplitude, and Fig.~5 j scores the direction of each effect. The outcome in Fig.~2 h is conducts, wiring-specific and the registered criterion together. The gap-junction row adds one connection per side by hand, from each giant fiber onto the tergotrochanteral motor neuron that receives its chemical synapses. Each was run at 13 (the chemical count of the two sides together), 52, 104, 208 and 416 synapse equivalents, and the two are the only hand-placed connections in the paper. A pathway's share of its target's input is its synapses onto the target over the target's total input synapses. MN9 and the giant fiber are each a pair of cells, and the input is counted over the pair. The taste receptor neurons make no synapse onto MN9. For the taste rows the share is that of the relay neurons that receive from the receptor neurons and synapse onto MN9 (38 neurons with 349 synapses for labellar sugar). In the gap-junction row the added equivalents are counted with the chemical synapses, in the pathway and in the total, and the share is the mean over the two motor neurons.

\subsection*{The odor probe}
The odor code (Fig.~3) was measured outside the body on the untrained network. The antennal receptor neurons were driven at 80 Hz for 0.5 s, then silenced. Before each presentation the network rested in silence for 1 s; an odor was then presented for 1 s by driving the receptor neurons of its five glomeruli at 150 Hz, and every neuron's spikes were counted. The main pair is A = D, DA1, DA2, DA3 and DA4l against B = DA4m, DC1, DC2, DC3 and DC4; three further pairs continue through the alphabetical list of glomeruli, from DL1 to VL1. For generalization (Fig.~3 h), A = DL5 and DM1 to DM4, and B shares $k$ of them, its other glomeruli drawn from DM5, DM6, DP1l, DP1m and V. A layer's active fraction, the quantity on the figure axes, is the fraction of its neurons with at least one spike during the presentation, and the text gives it as a percentage, the share of the layer's neurons active; the overlap of two odors is the fraction shared between the 5\% most active Kenyon cells of each; the MBON output is the total of MBON spikes in the presentation, and an output under 5 spikes before training is treated as unmeasurable.

The matched variant scales the output synapses of the 105 antennal-lobe local neurons predicted cholinergic by a gain $g$. The sweep used $g$ = 1, 0.5, 0.25, 0.15, 0.1, 0.075, 0.05 and 0; the variant used through the paper takes the largest gain at which the share of Kenyon cells active lies in 3 to 15\% with an overlap of at most 0.5, a rule fixed before the sweep ran, which gives $g = 0.15$. Training on odor A depresses each Kenyon-cell-to-MBON synapse by a fraction 0.5 (or 1.0) of its Kenyon cell's response to A normalized to the largest response, after which both odors are presented again; the depression of an odor is 1 minus the ratio of its MBON output after training to before. Silencing a type sets its output synapses to zero. The contingency test drives every receptor neuron at 0, 5, 40 or 120 Hz with no odor, after the warm-up or from silence; the breadth test builds odors of 5, 10 or 20 glomeruli; the drive test presents the odors at 0.5 and 0.25 of the 150 Hz rate. Six seeds per condition went to the gain sweep and the contingency test, whose results set the matched gain and the bistability, and three to the other tests; the seed sets only the Poisson input, and the seed-to-seed standard deviation of the share of Kenyon cells active is 0.03 percentage points as read and 0.17 at the matched gain.

One run tested the sign of the APL and was not pre-registered. The two APL cells, the typed right cell and the unlabeled left one, were signed inhibitory, as the table's verified column has them, and the odor code was measured as above, with and without training on odor A, at gains of 1, 0.5, 0.25, 0.15 and 0.05 on the cholinergic local neurons' output, three seeds each. Two further variants silenced the two cells, and gave every neuron with a single verified transmitter the sign of that transmitter, the unlabeled APL cell taking its twin's. The two cells, signed inhibitory, lower the share of Kenyon cells active, as read, from 68\% to 42\%, short of the fly's range, and applying every verified sign in the table gives the same 42\%. The verified column disagrees with the prediction's sign for 2,732 of the 60,448 neurons whose entry names one transmitter, 1.5\% of all synapses, 82\% of them in the optic lobe.

\subsection*{The body and its coupling to the network}
The body is NeuroMechFly\cite{LobatoRios2022,WangChen2024}, a MuJoCo model of the fly, stepped at 0.1 ms with leg adhesion on and contact sensors on the tibia and the five tarsal segments of every leg. Walking uses flygym's hybrid controller (a central-pattern-generator network with its retraction and stumbling reflexes), changed only to take six amplitudes, one per leg.

Every 50 ms the loop runs once. The arena's odor intensity at the two sensors on each side of the head, antenna and maxillary palp, is combined as a weighted mean (weights 9 and 1); the asymmetry $a = (L - R)/\text{mean}$, clipped to $\pm 1$, sets the receptor drive of the left and right antennae to $120 + 80a$ and $120 - 80a$ Hz. From each eye's ommatidia, 1 minus the mean brightness over the upper 45\% of rows is the looming signal; each side's LC and LPLC neurons receive 250 times this value, capped at 150 Hz. The network then runs for 50 ms, and the spike counts of each leg-motor pool, of the MBONs and of the left and right DNa02 are read.

The pool counts are smoothed by an exponential moving average with weight 0.15, and each leg's amplitude is $1 + 0.8\,(c - b)/s$, where $c$ is the smoothed count, $b$ the pool's baseline and $s$ the mean baseline over pools, both measured over the last 17 of 25 calibration steps at 120 Hz receptor drive before the episode and floored at 1; amplitudes are clipped to between 0.3 and 1.45. Steering scales the three legs of one side. The odor term is $e = \tanh(\beta^2)\,\mathrm{sign}(\beta)$ with $\beta = -500\,(L - R)/\text{mean}$. The valence $v$ is the spike count of the approach-driving MBONs minus that of the avoidance-driving MBONs, divided by a fixed norm and clipped to between $-1.5$ and 1.5. When $v$ is positive the legs on the side of the stronger odor are scaled by $1 - 0.8\,|e|\,v$; when it is negative the legs on the other side are scaled by $1 - 0.8\,|e|\,|v|$, and the fly turns away. Each MBON type drives approach or avoidance according to its transmitter, after the finding that activating a glutamatergic MBON repels flies and activating a GABAergic or cholinergic one attracts them\cite{Aso2014b}. The transmitter is the table's verified entry where the type has one, else the published one\cite{LiFeng2020} (MBON26, 27, 30 and 32), else the majority of the per-neuron predictions. Of the 104 MBONs read, 75 drive approach, 25 avoidance and 4 are unsigned (MBON24 and MBON29; Supplementary Table~5). The norm, 37.35 as read and 25.22 for the matched variant, is the intact network's mean balance over the calibration steps (53.78 and 36.32 spikes per step) divided by 1.44, the intact fly's untrained valence under the summed readout, so that the turn rule has the same scale under both readouts. The summed readout reads $v$ as the summed MBON count divided by 107, clipped to between 0 and 1.5; its results are reported in the Supplementary Discussion and Supplementary Tables~6 and 7. When the larger looming signal exceeds 0.05, a ramp $r$ rises linearly to 1 at 0.28. The odor term is multiplied by $1 - r$, and the legs opposite the stronger looming signal (or, when the two DNa02 counts differ, opposite the more active DNa02) are scaled by $1 - 2.2\,r$. Side scalings are clipped to between 0.3 and 1.

Arenas are built on flygym's odor arena\cite{WangChen2024}, with one odor whose intensity falls with the inverse square of the distance from each source; a source is consumed and switched off when the fly's center comes within 2.5 mm of it (4 mm in the learning arena). Walls are boxes 1.2 mm thick and 3 mm high. The fly starts at the origin facing $+x$. The layouts are ours. The nine arenas are an open field, a single detour, two-, three- and four-wall slaloms, a wide gate, two one-sided walls, a long S-curve and a dense slalom, with 32 sources in all and a 20 s cap. The forage arena of Fig.~5 has three sources at (26, $-6$), (28, $-9$) and (24, $-5$) mm behind a wall at $x = 15$ mm spanning $y = -4$ to 6 mm, and a 16 s cap. The learning arena has one source, at (22, 11) mm for the model as read, and 6 s episodes. A seed sets the random generators of the Poisson input and of the physics reset and, in rewired runs, the permutation.

\subsection*{The nine-arena battery}
Each of 20 seeds ran every arena under each of three wirings (intact, rewired and mushroom-body-lesioned), 180 runs per condition. The measures per run are the sources reached, the straightness (net displacement over path length), the median speed over 50 ms steps and the fraction of steps with speed above 1 mm/s. The chance level is a correlated random walk through the same nine geometries, with a heading perturbed every step, the intact fly's own speed (the mean over its runs of the median speed), reflection at walls, the 2.5 mm capture radius, the 20 s cap and 20 seeds. Two reference controllers with the same speed, capture radius, cap and seeds are given the position of every source and step straight toward the nearest one that remains; when a wall blocks a step, one stays where it is and the other steps along the wall (Supplementary Table~1). The recordings of Fig.~4 e, l and m are every neuron's spike count in every 50 ms step of one run per condition (arena i, seed 0).

\subsection*{Perturbations in the body}
The forage-arena conditions (Fig.~5 b to d) are intact; the mushroom body lesioned; rewired; the lesioned and the rewired fly with the valence pinned at 1.44, the intact fly's mean under the summed readout (1.41 with the balance); the lesioned fly pinned at 0 and at 0.7 and the intact fly pinned at 0; a random lesion of 4,608 neurons, the size of the mushroom-body set; and the intact fly with the looming signal set to zero. Each has 20 completed runs, taken in seed order; runs that ended in a physics fault are excluded and counted, and extra seeds were run to fill the 20, under a rule written before they ran. The six measures are the sources reached, the mean over sources of the closest approach, the mean DNa02 count per step, the number of steps within 0.5 mm of the wall, the path length and the leg motor rate. The leg motor rate is the spikes per second of one leg's motor pool, summed over its 63 to 70 neurons and averaged over the six pools and the steps of the run. The intact fly's 154 Hz is about 2.4 Hz per neuron.

Motor replay (Fig.~5 e, g) ran each arena and seed three times. A closed-loop run from the standard start recorded its six amplitudes at every step. A second closed-loop run started with the heading rotated by 30$^{\circ}$. An open-loop run from the same rotated start fed the recorded amplitudes to the body step by step while the network's output was ignored.

Descending-neuron drive (Fig.~5 f, j) took place in an open arena with the receptor neurons at 120 Hz on both sides (also 40 and 0 Hz for j). Each side scaling was fixed at 1 in these runs, which switched off every turn rule. Vision was not simulated, and the looming input stayed at zero. After 1 s the target received 600 Hz for 2 s, then 1 s more without it. The targets were DNa01 and DNa02 on each side, DNa02 on both sides, DNp09 on the left, the connectome's threat-response cluster on both sides, DNg12 and DNp01 on the left as non-locomotor controls, and no target as the baseline; 20 seeds each intact, and for the rewired rows three permutations with seven seeds each. Speed is path length over time, in the drive window against the settling window (0.2 to 1 s), and the net turn is the change in heading over the drive window. In Fig.~5 j a cell carries the animal's sign when its effect has the direction measured in the animal and exceeds twice its s.e.m. (and 0.01 Hz in the silent network, 0.10 Hz on the background). An effect that exceeds them in the other direction is the opposite sign, and a smaller one is no effect. The animal's sign for DNg12 and DNp01 is no change in walking, and a cell carries it when the effect stays within that threshold. The threshold is looser than the three control standard deviations registered for Fig.~2, which the 2 Hz effect of DNa02 misses from either side. The body rows score the net turn for a target that turns the animal (DNa01, one-sided DNa02 and DNp09), the speed change for bilateral DNa02 and the threat cluster, and both for DNg12 and DNp01. Fig.~5 f asks more of a target whose sign has two parts, and fills its marker when both are carried. DNp09 turns as the animal does and slows where the animal speeds up. Its turn carries the sign in j at 120 Hz, and its speed leaves the marker in f open. The two wiring rows read the leg-motor drive of Fig.~2 as the body would, with a fall on the driven side taken as a turn toward it and a rise in the total drive as faster walking. Six cells of the two wiring rows come from a further sweep, run as those of Fig.~2 were with nine seeds, rewired twins and criteria registered before it ran. They are DNa02 on the right and DNp01 at 2 Hz, and bilateral DNa02 and the threat cluster in the silent network and at 2 Hz.

Graded removal (Fig.~5 h, i) zeroes every synapse of a fraction of the candidate neurons, all simulated neurons except the receptor, LC and LPLC, leg-motor, MBON and DNa02 sets that the loop drives or reads. Random removal draws the set with the seed, at 5 to 90\%; targeted removal takes the neurons with the most synapses at 1 to 20\%, over all candidates or over the central-brain intrinsic neurons only; 20 seeds each. A run that ended in a physics fault is excluded and counted, as in the forage arena, and no further seeds were run. A point is the mean over the runs that completed, 20 in 14 of the 21 conditions, 17 to 19 in six and 9 at 1\% over all candidates. Every run that faulted had reached no source. Counting those runs as zeros would lower three means, 2.53 to 2.40 at 5\% random removal, 0.18 to 0.15 at 70\% and 0.78 to 0.70 at 1\% within the central brain, and leave the others as they are.

\subsection*{Learning in the loop}
Of the 15,523 connections from Kenyon cells onto MBONs, the plastic ones are the 8,377 onto the MBONs of six of the seven compartments that the PPL1 dopamine neurons innervate, the compartments of aversive learning (types MBON11 to 14, 16 to 20, 23, 25, 28, 30 and 32 to 35, 50 cells, after the hemibrain compartment map\cite{LiFeng2020}; mean weight 0.513 mV). The six are $\gamma1$, $\gamma2$, $\alpha'2$, $\alpha'3$, $\alpha2$ and $\alpha3$. The other 7,146 connections keep their weight, among them those onto the three types confined to the seventh compartment, $\alpha'1$ (MBON15, MBON15-like and MBON31, 10 cells). At every 50 ms step each plastic connection's eligibility trace decays with a 200 ms time constant and gains its Kenyon cell's spikes in the step. While the reinforcement signal is on, the connection accumulates a depression of $\eta$ times its eligibility times its MBON's spikes in the step; at the end of the episode the accumulated depression is subtracted from the weight, which is clipped to between 0 and its initial value. In the fly an odor paired with dopamine depresses synapses of this class even when the MBON's spikes are blocked\cite{Hige2015}, and the rule here differs from the fly's in that factor. A control without that factor ran the paired and unpaired conditions below with 20 seeds in both models (Supplementary Table~9). Its predictions were written down before it ran, and all of them held. In the matched variant the control's unpaired condition leaves the mean weight of the plastic connections at 0.455 mV and the valence at 0.90, against 0.508 mV and 1.38 with the factor. All 20 of those flies still reach the source, and none of the 20 paired flies does. The registered tests of the variant, defined below, read associative and expressed as they do with the factor, and the network as read is not associative under either rule. Under the summed readout with all 15,523 connections plastic (Supplementary Discussion) the learning lowers the valence in both conditions, but not below a floor set by MBONs the Kenyon cells do not drive, and 19 of 20 trained flies still reach the source. Two further arms change one of the two at a time (Supplementary Fig.~4; Supplementary Table~7). One is the balance readout with all 15,523 connections plastic, and the other is the summed count (divided by 107 as read and by 81 in the matched variant, as in the Supplementary Discussion) with the 8,377. Each ran the three conditions below with 20 seeds in both models, and its predictions were written down before it ran.

A run is a naive test episode, six training episodes and a post-test episode, each 6 s from the same start, with the source switched on at the start and off at contact. In the paired condition the reinforcement is on throughout each training episode while the fly forages in the odor. In the unpaired condition the reinforcement is on but the odor is withheld, both antennae receiving the 5 Hz the arena supplies without an odor. In the no-plasticity condition $\eta = 0$. The main arm used $\eta = 0.02$ and 20 seeds per condition; the learning-rate ladder ran the paired condition at eleven values of $\eta$ from $10^{-5}$ to 0.08, 20 seeds each. The measures are the mean weight of the plastic connections (0.513 to 0.138 mV after pairing as read; a mean over all 15,523 connections, which includes those that cannot change, reads 0.500 to 0.298 mV), the valence averaged over an episode, the closest approach to the source in the post test, and the devaluation, the fall from the naive to the post test in the fraction of the start distance closed. The matched variant (Fig.~6 i, k) is the $g = 0.15$ network of Fig.~3. Its balance norm, 25.22, was set as above from its own intact runs, and its source was placed at (23, 8) mm by a rule fixed before the runs, which transfers the as-read source's distance from the start, side and margin to the variant's own path with the valence forced to zero. The registered tests of the variant are on the behavior. The depression is associative when the paired and unpaired flies' devaluation differ ($P < 0.01$, Cliff's $\delta \ge 0.5$) with the unpaired valence at or above 0.7, and expressed when more than half of the trained flies miss the source by more than 4 mm while the unpaired and unplastic flies reach it.

\subsection*{Statistics and reproducibility}
The unit of replication is the seed, one simulated fly with its own input and initial conditions; arenas are repeated measures on a fly. All tests are two-sided. The replay result (Fig.~5 g) is $W = 0$, $P = 1.9 \times 10^{-6}$ for 20 flies, a paired Wilcoxon signed-rank test on each fly's mean fraction of sources reached over the nine arenas, computed exactly (SciPy 1.17.1, exact method). The same exact test on the rotated closed loop against the unrotated record gives $P = 0.13$. Group comparisons (Fig.~2 g, Fig.~6 f, g, i, Supplementary Tables~7 and 9, and the rewired against the lesioned fly in Fig.~4 f) use the Mann--Whitney $U$ test with Cliff's $\delta$ (the probability that a value from one group exceeds one from the other, minus the reverse) as the effect size, and 95\% intervals from 5,000 percentile-bootstrap resamples, reported in the Results where the two groups overlap (at $\delta = \pm 1$ every resample gives the same value). The Mann--Whitney $P$ values come from the normal approximation with continuity and tie corrections, SciPy's method when both groups have more than eight values. The approximation gives $P = 6.8 \times 10^{-8}$ where every value of one group lies above every value of the other, 20 against 20, and the exact test gives $1.5 \times 10^{-11}$. The lesion is compared with the random lesion of the same size (Fig.~5 d) by Fisher's exact test on the number of runs that reach no source. The power of a 20-versus-20 Mann--Whitney test at $\alpha = 0.05$ against a normal shift whose $\delta$ is 0.2, 0.35, 0.5 or 0.65 is 0.19, 0.49, 0.83 and 0.97 (2,000 simulations each). The correlation of Fig.~3 i is Spearman's $\rho$ over 11 types ($\rho = 0.78$, $P = 0.0045$ from the $t$ distribution); without the largest type it is 0.71 ($P = 0.022$), and leaving out any one type gives 0.71 to 0.88. No correction for multiple comparisons was applied; a Bonferroni correction across the six learning tests changes no conclusion.

Every sweep's cells, metrics, thresholds and predictions were written down before it was launched, except the exploratory APL run above. Simulations ran in one container (Ubuntu 22.04, Python 3.10.12, Brian 2 2.7.1 with Cython 0.29.37, flygym 1.2.1, MuJoCo 3.2.7\cite{Todorov2012}, NumPy 2.0.2\cite{Harris2020}, SciPy 1.15.3\cite{Virtanen2020}), one run per 2-core, 12 GiB pod on the National Research Platform's Nautilus cluster; figures and statistics were built locally (Python 3.11, SciPy 1.17.1).

\bibliographystyle{naturemag}
\bibliography{main}

\section*{Acknowledgements}
We thank the BANC-FlyWire Consortium and the wider \textit{Drosophila} community for proofreading and annotating the
BANC connectome, Bates et al. for the connectome and its transmitter predictions, and the FlyWire team for the tools
through which it is distributed; FlyWire is supported by the NIH BRAIN Initiative (RF1 MH117815, RF1 MH129268 and
U24 NS126935 to M. Murthy and H. S. Seung). The simulations ran on the National Research Platform, which is supported in
part by National Science Foundation awards CNS-1730158, ACI-1540112, ACI-1541349, OAC-1826967, OAC-2112167, CNS-2100237
and CNS-2120019.

During the preparation of this work, the authors used the Claude large language model (Fable 5.1, Fable 5.2, Opus 5) in order to
generate portions of the source code, analysis figures, and in editing the text. After using these tools, the authors
reviewed and edited the content as needed and take full responsibility for the content of the publication.

\section*{Funding}
M.T. discloses support for the research of this work from Schmidt Futures (SF857), the National Human Genome Research
Institute (RM1HG011543), the National Science Foundation (2134955, 2034037 and 2515389), the California Institute for
Regenerative Medicine (DISC4-16285) and the University of California Office of the President (M25PR9045). D.H. discloses
support from Schmidt Futures (SF857), the National Human Genome Research Institute (RM1HG011543), the National Science
Foundation (2134955 and 2515389), the National Institute of Mental Health (U24MH132628) and the National Institute of
Neurological Disorders and Stroke (U24NS146314). M.A.M.-R. discloses support from the National Science Foundation
(2515389), the National Institute of Mental Health (U24MH132628), the National Institute of Neurological Disorders and
Stroke (U24NS146314), the California Institute for Regenerative Medicine (DISC4-16285, DISC4-16337 and DISC4-19334), the
University of California Office of the President (M25PR9045) and the Brain and Behavior Research Foundation (33184).
J.G.-F. discloses support from the QB3 Graduate Innovators Fellowship. J.R.M., J.G. and J.K.E. declare no relevant
funding. The funders took no part in the design of the study, the analysis, the decision to publish or the writing of the
manuscript. The content is the authors' responsibility alone and does not necessarily represent the official views of the
National Institutes of Health, the National Science Foundation, the University of California, the California Institute
for Regenerative Medicine or any other federal or state agency.

\section*{Author contributions}
J.R.M. designed the study and performed the experiments. M.T. supervised the work. J.G.-F., J.G., J.K.E., M.A.M.-R.,
D.H. and M.T. contributed to study design and interpretation. M.T., D.H. and M.A.M.-R. acquired funding. J.R.M. wrote
the manuscript, with contributions from all authors. All authors read and approved the manuscript.

\section*{Competing interests}
The authors declare no competing interests.

\section*{Materials and correspondence}
Correspondence and requests for materials should be addressed to J.R.M. or M.T.

\section*{Figure legends}
\noindent\textbf{Fig.~1.} \capFigOne\par\medskip
\noindent\textbf{Fig.~2.} \capFigTwo\par\medskip
\noindent\textbf{Fig.~3.} \capFigThree\par\medskip
\noindent\textbf{Fig.~4.} \capFigFour\par\medskip
\noindent\textbf{Fig.~5.} \capFigFive\par\medskip
\noindent\textbf{Fig.~6.} \capFigSix\par\medskip

\clearpage
\mainfigurelandscape{width=\linewidth,height=0.93\textheight,keepaspectratio}{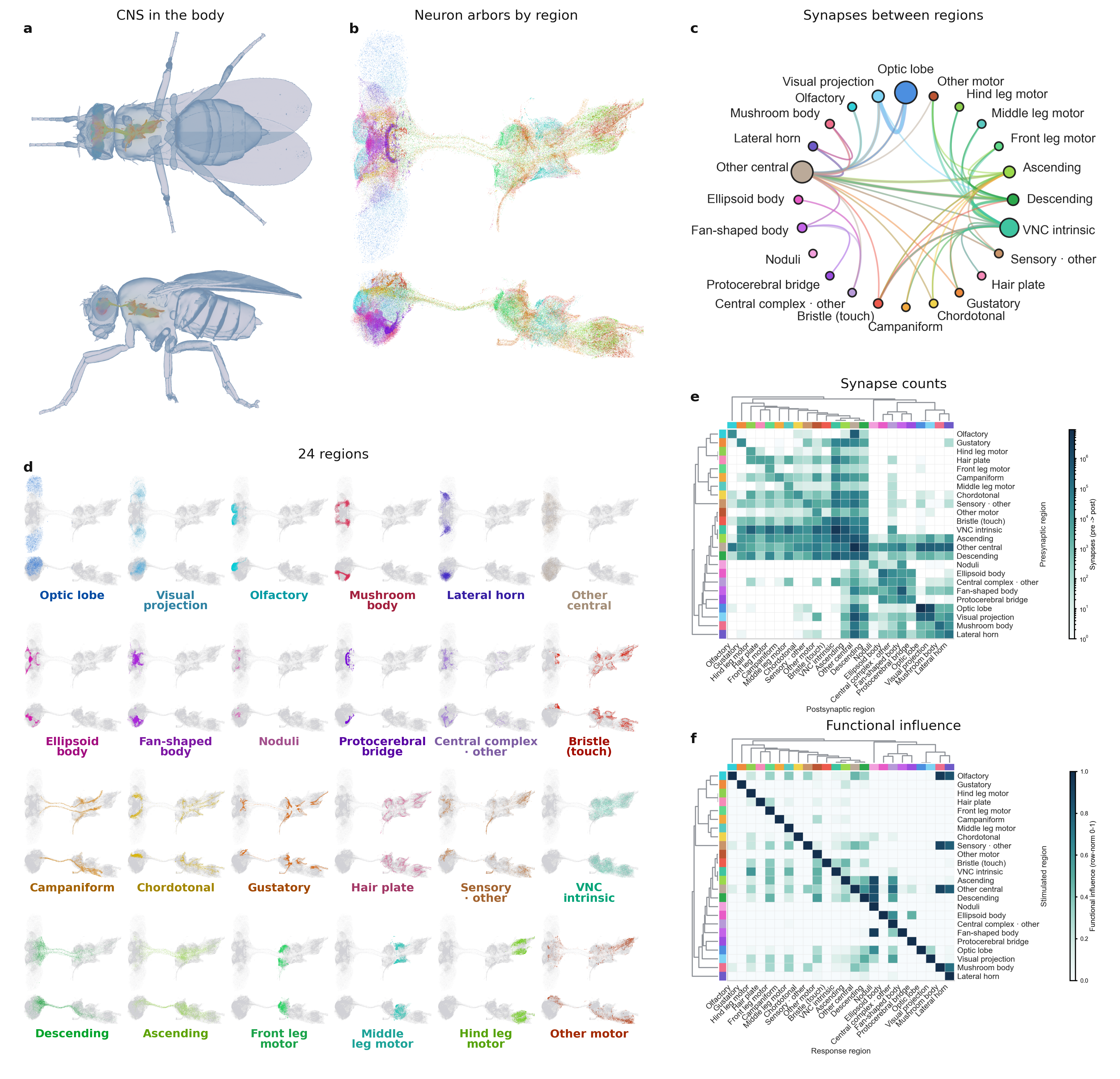}{1}
\clearpage
\mainfigure{width=\linewidth,height=0.93\textheight,keepaspectratio}{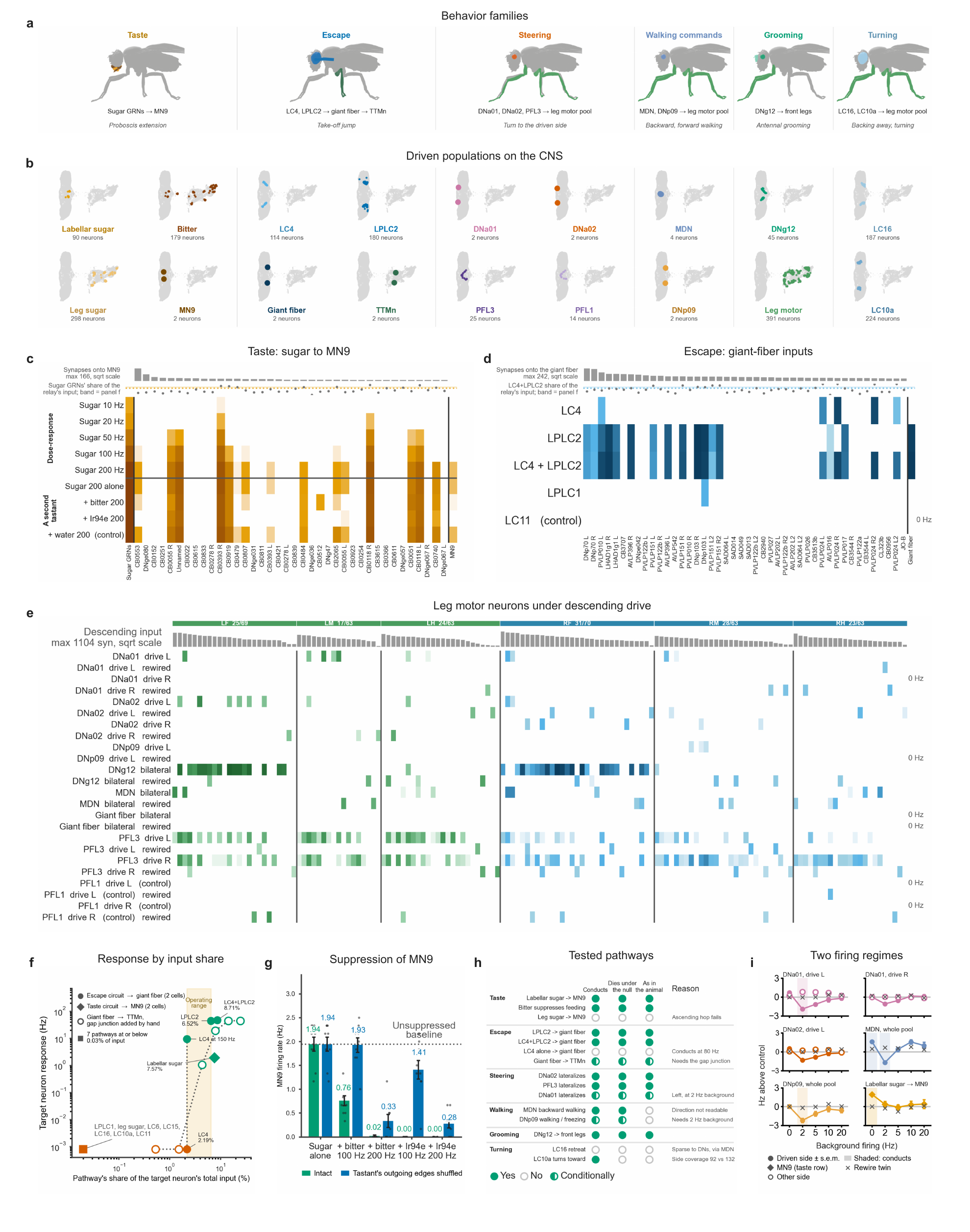}{2}
\clearpage
\mainfigurelandscape{width=\linewidth,height=0.93\textheight,keepaspectratio}{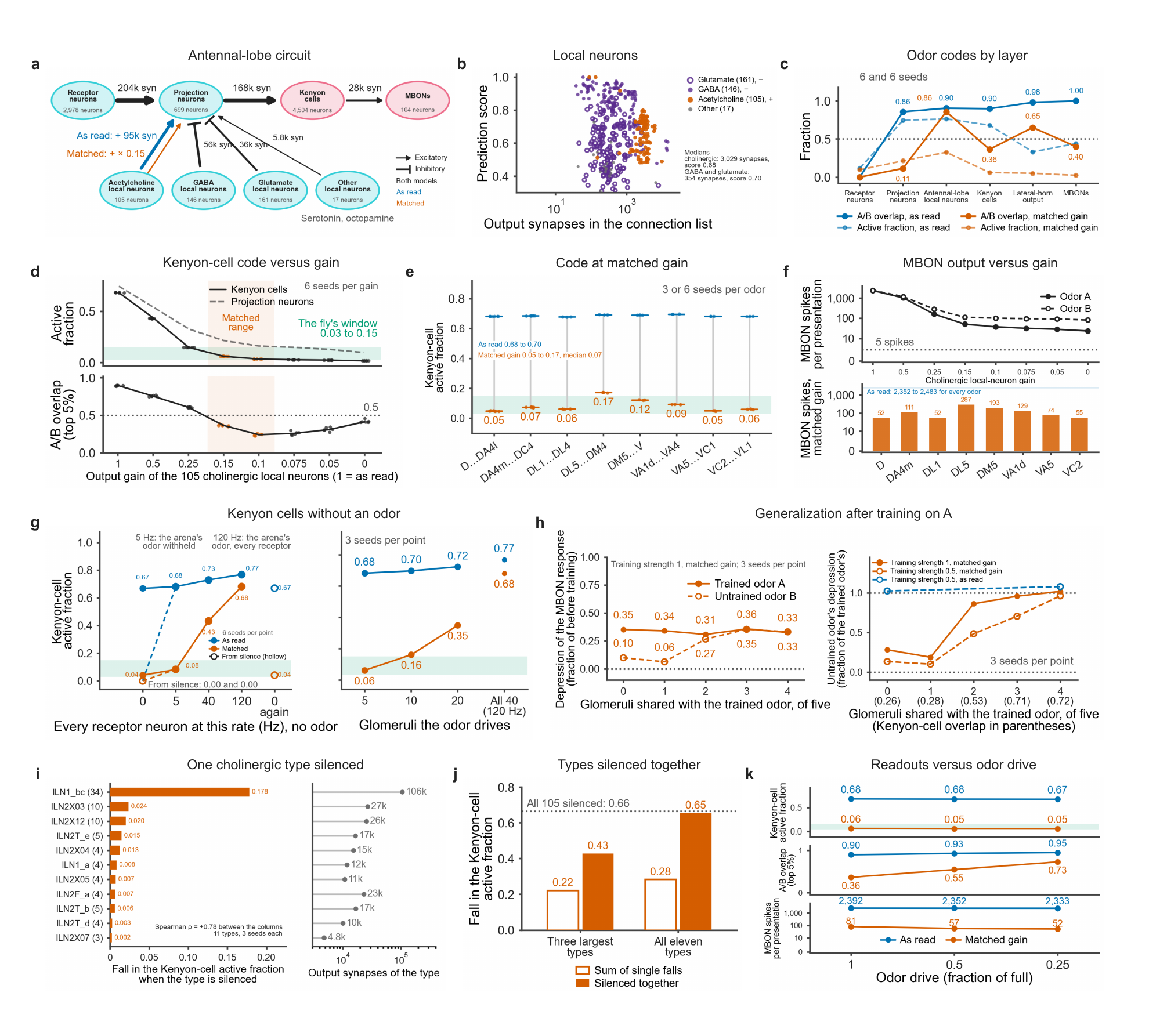}{3}
\clearpage
\mainfigure{width=\linewidth,height=0.93\textheight,keepaspectratio}{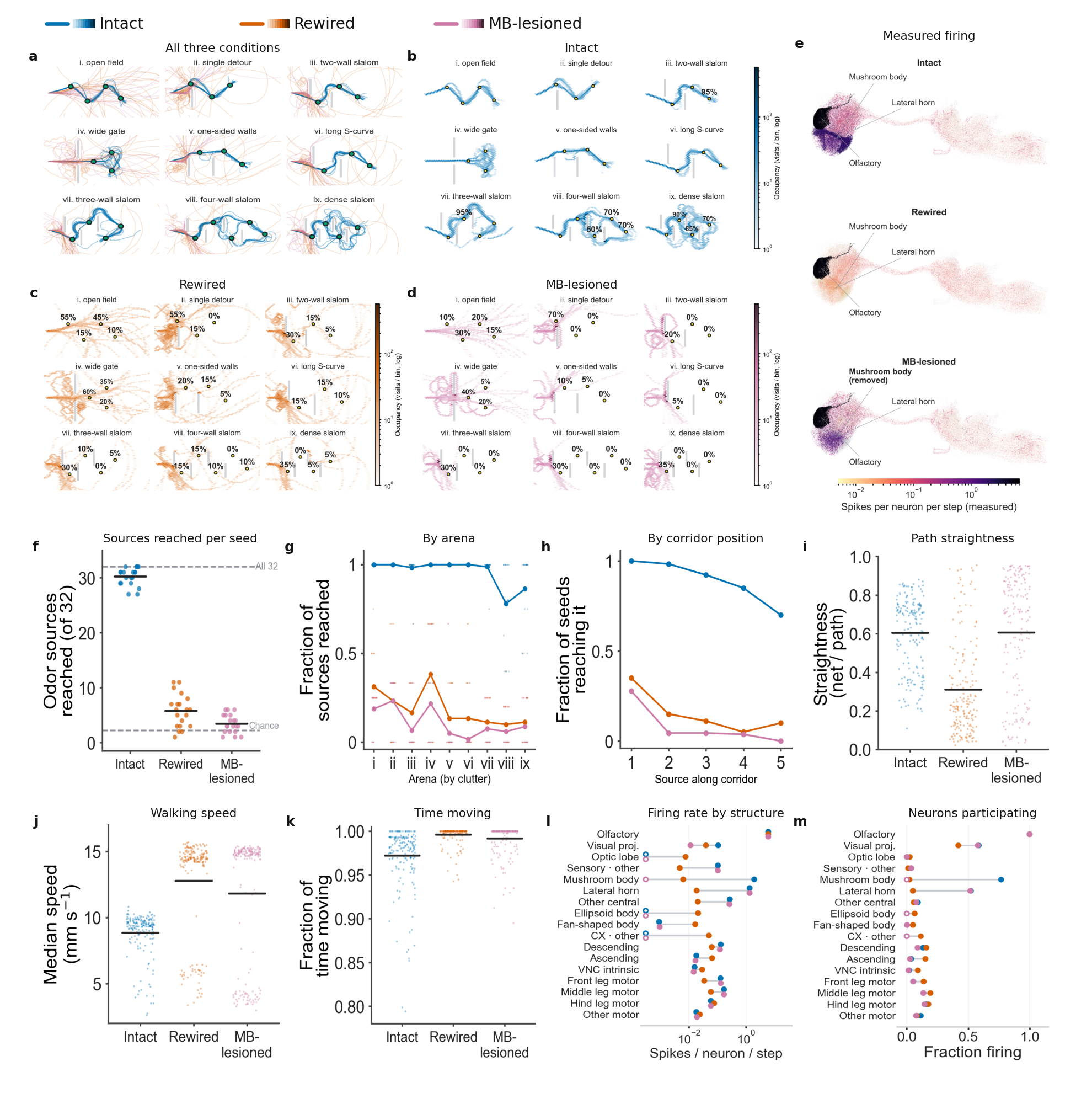}{4}
\clearpage
\mainfigurelandscape{width=\linewidth,height=0.93\textheight,keepaspectratio}{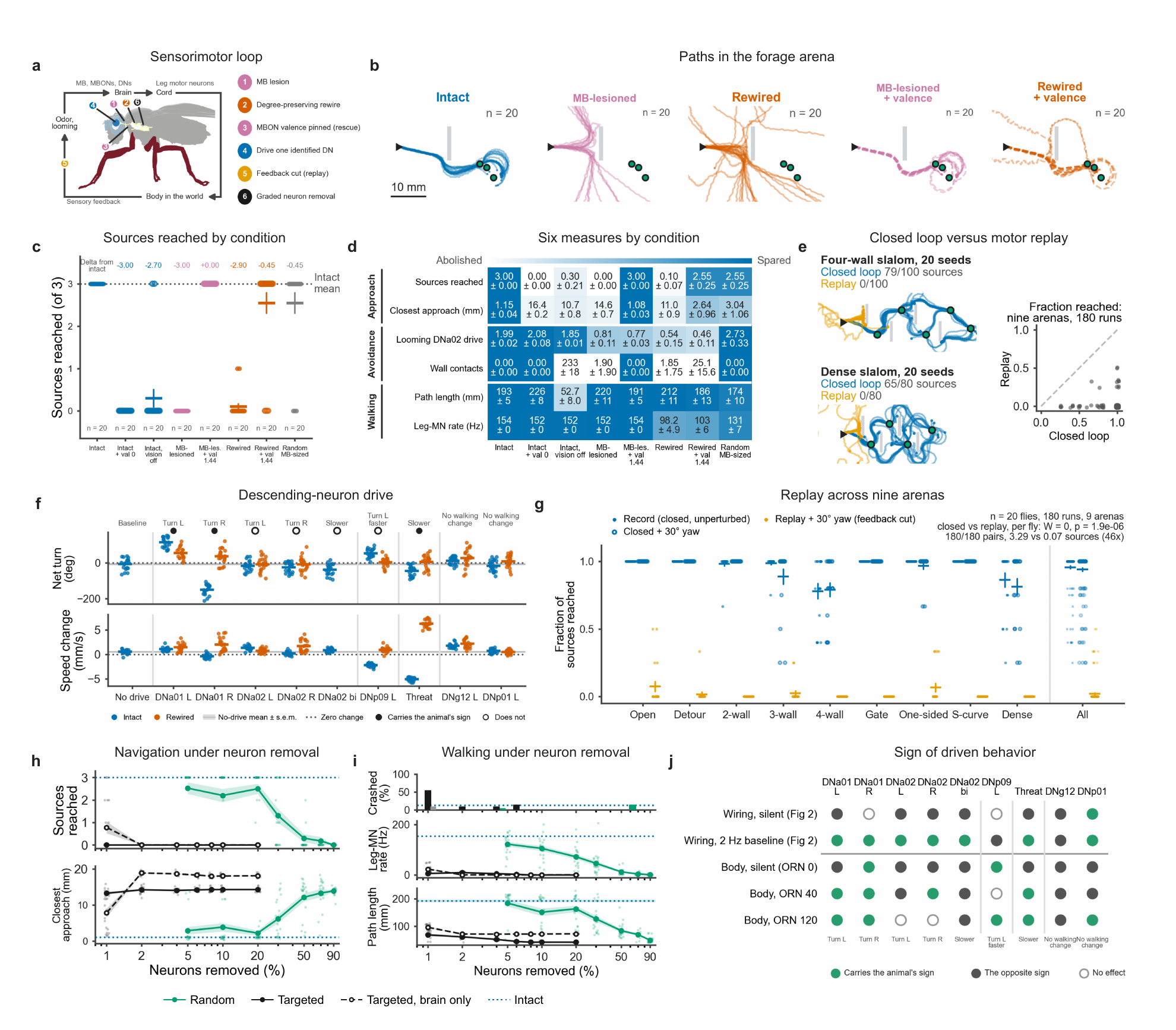}{5}
\clearpage
\mainfigurelandscape{width=\linewidth,height=0.93\textheight,keepaspectratio}{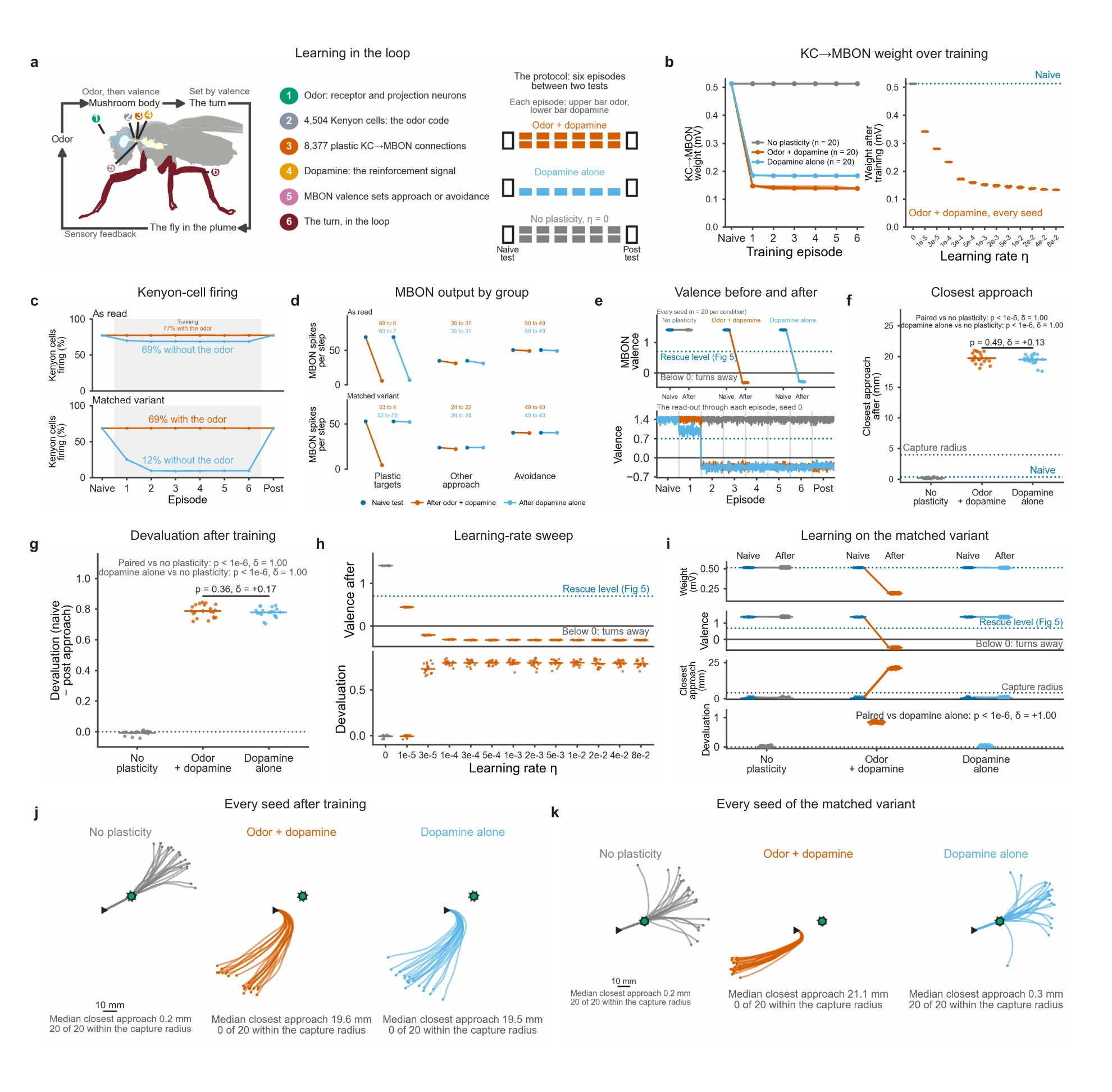}{6}

\clearpage
\newgeometry{margin=2.0cm}
\begin{center}
{\Large\bfseries Supplementary Information}\par
\vspace{0.6em}
{\large Sensorimotor behavior from an untrained brain-and-nerve-cord connectome
\par}
\vspace{0.6em}
John R. Minnick, Jesus Gonzalez-Ferrer, Jinghui Geng, Jason K. Eshraghian, Mohammed A. Mostajo-Radji, David Haussler and Mircea Teodorescu
\end{center}

\noindent\textbf{Contents.} Supplementary Discussion, the learning experiment with the valence read as the summed
MBON count. Supplementary Fig. 1, a sensory pulse crossing the network, the odor and the looming pulse.
Supplementary Fig. 2, the order in which the wiring is traversed against the order in which the pulse reaches neurons.
Supplementary Fig. 3, the signed synapse matrix of the 24 regions beside the influence map.
Supplementary Fig. 4, the learned avoidance under the four combinations of valence readout and
reinforcement. Supplementary Table 1, sources reached in the nine arenas beside three reference
controllers. Supplementary Table 2, the conditions of the forage arena. Supplementary Table 3, descending-neuron drive
in the body at three receptor rates. Supplementary Table 4, the odor code against the gain of the cholinergic local
neurons. Supplementary Table 5, the sign of every MBON type under the balance readout. Supplementary Table 6, the
in-body results under the two readouts of the valence. Supplementary Table 7, the learning experiment under the four
combinations of valence readout and reinforcement. Supplementary Table 8, the 24 groups of Fig. 1 with their defining
labels and neuron counts. Supplementary Table 9, the learning experiment with and without the MBON's spikes in the
plasticity rule.

\subsection*{Supplementary Discussion. The learning experiment with the valence read as the summed MBON count}

The summed readout takes the valence as the spike count of all 104 MBONs in a 50 ms step divided by 107, clipped to between 0 and 1.5. The runs reported here let the reinforcement reach all 15,523 Kenyon-cell-to-MBON connections. Six training episodes
with the odor paired with the reinforcement took the mean weight from 0.500 to 0.128 mV and the valence from 1.44 to
0.24. The behavior did not change, with the trained fly's closest approach to the source in the post test at a median 0.23 mm against 0.20 mm before training. The trained flies still reached the source, 19 of 20, as did 20 of 20 without plasticity ($P = 0.23$ and 0.47 against no plasticity on closest approach and devaluation, two-sided Mann--Whitney $U$ tests). The valence after training fell to between 0.24 and 0.25 at learning rates from $\eta = 3 \times 10^{-4}$ upward and no further, and 18 to 20 of 20 flies reached the source at every rate.

The floor was MBON output that the plasticity cannot lower. The MBON spikes per step fell from 155 to 26 with training, and ten MBONs carried 87\% of the 26. Six of the ten take at most a quarter of their input from Kenyon cells
(MBON19, two MBON26 cells, MBON04 and two MBON31 cells). The two MBON05 cells take a third and a half and are driven
besides by the APL and the glutamatergic MBON01. The two that are mostly driven by Kenyon cells (MBON11 and MBON06)
are the two that fell most. A valence pinned at 0 abolishes approach in the forage arena and one pinned at 0.7 keeps it (Fig.~5). The switch of the learning arena lay below 0.24, with 18 to 20 of 20 flies reaching the source at a valence between 0.24 and 0.28. In the matched variant the depression was associative at the synapse. Pairing took the
weight to 0.182 mV and the valence to 0.25, while reinforcement alone left the weight at 0.495 mV and the valence at
1.42. The matched variant's trained fly still reached the source (19 of 20), because its valence fell to the same floor.

A sum over every MBON cannot fall below the output of the MBONs that learning does not touch, and it cannot change
sign. MBON types in the fly differ in the behavior they drive, and learning shifts the balance between those that drive approach and those that drive avoidance (Methods). The main text reads the valence as that balance and confines
the reinforcement to the compartments of aversive learning (Methods; Supplementary Table 5). The balance readout, its predictions and a pilot of 150 runs were written down before the in-body runs of Figs 4 to 6 were made. Supplementary Table 6 sets the two
readouts side by side. The lesion, the rescues, the pinned conditions and
the replay give the same results under both. Two other results differ most between the two readouts. The intact fly stops less often at walls under the balance readout (30.25 against 29.20 of 32 sources in the nine arenas, and all three sources of the forage arena in 20 against 16 of 20 seeds). The rewired fly reaches more sources under the balance readout (5.80 against 3.50 of 32). Its MBONs fire as much under both readouts, about 7.5 spikes per step, and the balance readout reads that as a valence of 0.11 where the summed readout reads 0.07, because 75 of the 104 MBONs drive approach.

The analysis of the main text differs from the summed readout's in two things: the readout, and which connections
the reinforcement reaches. Two further arms change one of them at a time, each with the three conditions, 20 flies
and both models of the learning experiment, and with predictions written down before the runs (Supplementary
Fig.~4; Supplementary Table~7). With the summed readout and the reinforcement confined to the aversive compartments, the valence after pairing is 0.83 (0.84 in the matched variant). It stays above the floor of 0.24 because the MBONs of the other compartments keep
their output, and all 20 flies reach the source in every condition. With the balance readout and the reinforcement in every compartment, the depression lowers the approach-driving and the avoidance-driving MBONs alike, to 13.7 and 12.5 spikes per step as read (10.0 and 10.5 in the matched variant). The valence after pairing is 0.03 ($-0.02$), and the trained fly then neither approaches nor avoids. It passes the source at a median 11.2 mm (10.2 mm), farther than every fly without plasticity and nearer than every trained fly of the main text's analysis ($P = 6.8 \times 10^{-8}$ for each comparison in each model). The source is still reached on the way by 2 of 20 flies (1 of 20). One registered prediction, that this count would lie between 2 and 12 as it had in three earlier runs with the valence held at zero, missed by one fly in the matched variant. Reinforcement alone does what pairing does in the network as read and changes nothing in the matched variant, in both arms as in the main text. The trained fly turns away only when the readout is signed and the reinforcement is confined to the
aversive compartments, where 41 of the 50 MBONs drive approach.

\begin{figure}[!htb]\centering
\includegraphics[width=\textwidth]{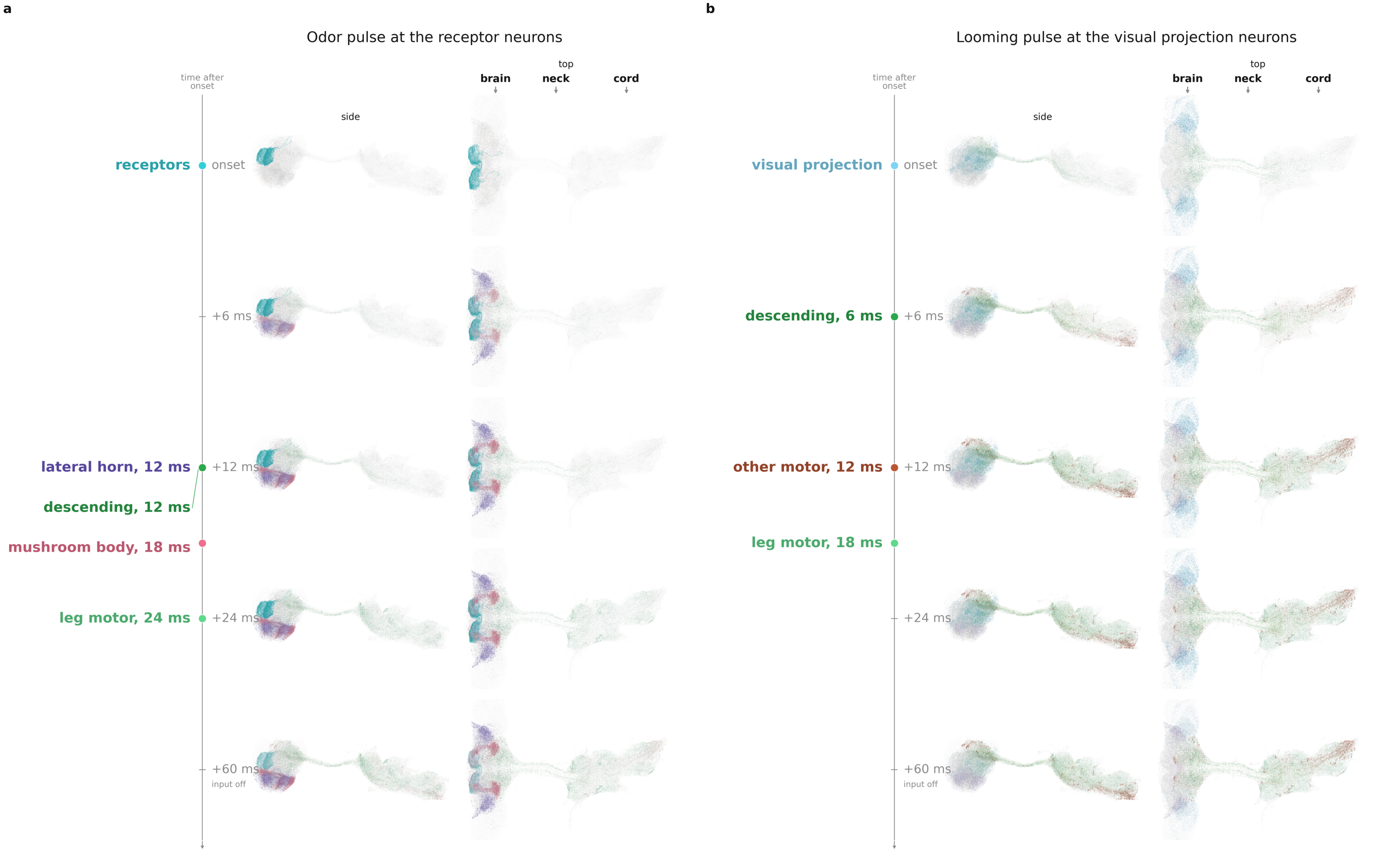}
\caption*{\textbf{Supplementary Fig. 1. A sensory pulse crosses the network from its input to the leg motor pools within 24 ms.}
Spiking activity after a 60 ms pulse at the receptor neurons (\textbf{a}, 400 Hz) or at the LC and LPLC visual projection
neurons (\textbf{b}, 300 Hz), in the network alone with no other input. Rows, the 6 ms bins starting 0, 6, 12, 24 and 60 ms
after the pulse's onset, on the brain from the side and on the whole nervous system from above; a region's color is its rate
in that bin, spikes per neuron per 6 ms, on a logarithmic scale from 0.001 to the input layer's peak (2.33 in \textbf{a}, 1.04
in \textbf{b}), over a gray outline of every region, in the colors of Fig.~1d. The time axis at the left carries each stage's
arrival, the end of the first bin in which the region's rate reaches a tenth of its mean over the pulse (the rule of the
Methods), with the region's name and the time. The input stops after 60 ms; the last row shows the brain still active after it.}
\end{figure}
\clearpage

\begin{figure}[!htb]\centering
\includegraphics[width=\textwidth]{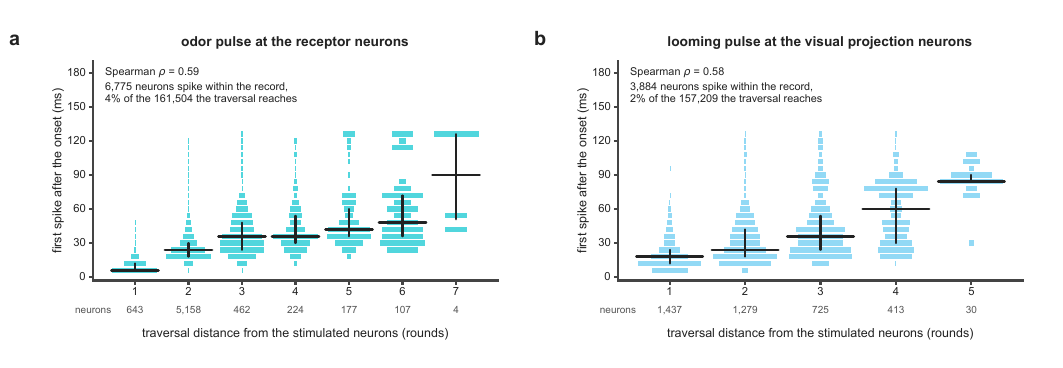}
\caption*{\textbf{Supplementary Fig. 2. The order in which the wiring is traversed from the stimulated neurons predicts the order
in which the pulse reaches neurons.} Every neuron that spikes within the record after the odor pulse (\textbf{a}, 6,775 neurons) or the
looming pulse (\textbf{b}, 3,884), its first spike after the onset, the end of the first 6 ms bin with a spike in the records of
Supplementary Fig.~1, against its traversal distance from the stimulated group in the wiring. The traversal is the information-flow
model of Schlegel et al. (Methods): the connection list is walked in rounds from the stimulated neurons, an untraversed neuron is
traversed with probability $\min(1, r/0.3)$, $r$ the share of its input synapses from traversed neurons, and the distance is the round,
averaged over ten seeded traversals and rounded. For each round, a bar at each 6 ms value whose length is the number of neurons with
that first spike, the round scaled to its own peak, in the stimulated region's color of Fig.~1; black bar, the round's median first
spike; vertical line, its quartiles; the count under each round, its neurons. Each panel gives Spearman's $\rho$ over its neurons and
the share of the neurons the traversal reaches that spike within the record: the traversal reaches almost every neuron, the pulse in
the silent network 4\% and 2\% of them.}
\end{figure}
\clearpage

\begin{figure}[!htb]\centering
\includegraphics[width=\textwidth]{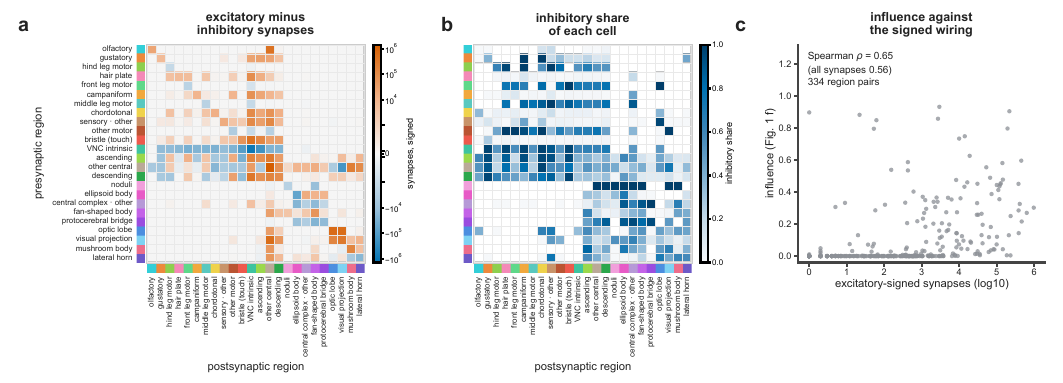}
\caption*{\textbf{Supplementary Fig. 3. The signed wiring between the 24 regions, and the influence map against it.} \textbf{a}, The
synapse counts of Fig.~1e split by the sign the model gives the presynaptic neuron, excitatory minus inhibitory, on a symmetric log
scale (vermillion, an excitatory excess; blue, an inhibitory excess); 42.5\% of all synapses carry an inhibitory sign. \textbf{b}, The
inhibitory share of each cell's synapses. Rows and columns of \textbf{a} and \textbf{b} are in the order of Fig.~1e and f, with the
row labels on \textbf{a} and each region's color, as in Fig.~1, along the edges. The rows of Fig.~1f that evoke nothing elsewhere are
the rows of \textbf{b} whose output is mostly inhibitory-signed, or too small to drive a region (Results). \textbf{c}, Per region pair
with any synapse (334), the influence of Fig.~1f against the excitatory-signed count; Spearman's $\rho = 0.65$, and 0.56 against the
count regardless of sign. The pairs with influence and no direct synapse are relayed routes, the receptors' onto the mushroom body
and the lateral horn among them.}
\end{figure}
\clearpage

\begin{figure}[!htb]\centering
\includegraphics[width=\textwidth]{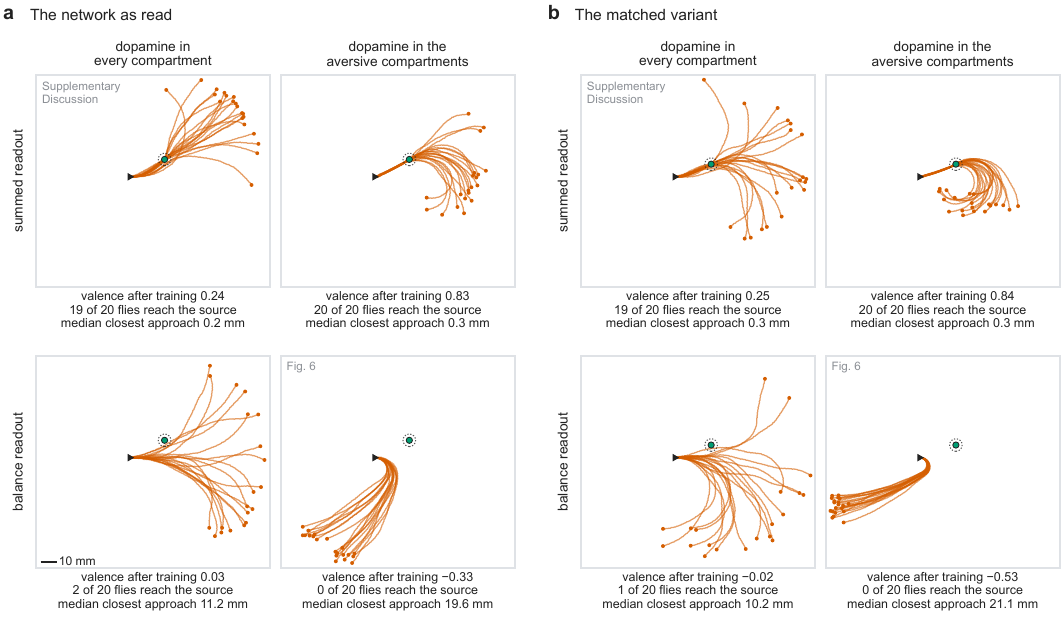}
\caption*{\textbf{Supplementary Fig. 4. The learned avoidance needs both the balance readout and reinforcement
confined to the aversive compartments.} The path of every fly in the post test after six episodes of odor paired
with the reinforcement (odor + dopamine; $n = 20$ flies per arena), under the four combinations of valence readout
(rows) and of the connections the reinforcement reaches (columns). \textbf{a}, The network as read. \textbf{b}, The
matched variant. Summed readout, the valence is the summed spike count of every MBON; balance readout, the count of the
approach-driving MBONs minus that of the avoidance-driving MBONs. Every compartment, the reinforcement reaches all
15,523 Kenyon-cell-to-MBON connections; aversive compartments, the 8,377 of six of the seven compartments that the PPL1
dopamine neurons innervate (Methods). The upper left arena of each panel is the summed readout with the reinforcement in every compartment
(Supplementary Discussion) and the lower right the analysis of the main text (the runs of Fig.~6 j and k); the other two
change one of the two at a time. Triangle, the start; green dot, the odor source, with its 4 mm capture radius dotted; a
dot ends each path. All eight arenas share one frame and one scale; bar, 10 mm. Under each arena: the valence in the post test
(mean over flies), the flies that reach the source, and the median closest approach to it. With the summed readout the trained flies reach the source wherever the reinforcement acts (19 or 20 of 20). With the balance readout and the reinforcement in
every compartment the valence goes to zero and the fly walks past the source, steering neither toward it nor away.
It turns away only with the balance readout and the reinforcement confined to the aversive compartments. Every
condition, with the tests, is in Supplementary Table~7.}
\end{figure}
\clearpage

\begin{table}[!htbp]\centering\small
\caption*{\textbf{Supplementary Table 1. Sources reached in the nine arenas, beside three reference controllers.} Mean sources reached over 20 seeds. Intact, rewired and lesioned are the three wirings of Fig.~4. The random walk is the chance level of Fig.~4f. The two informed controllers are given the position of every source and head straight for the nearest one that remains, at the intact fly's median speed and with its capture radius, start and time limit; at a wall one stops and the other slides along it.}
\resizebox{\ifdim\width>\textwidth\textwidth\else\width\fi}{!}{%
\begin{tabular}{llrrrrrrrr}
\toprule
Arena & Layout & Walls & Sources & Intact & Rewired & Lesioned & Random walk & \shortstack{Informed,\\stops at walls} & \shortstack{Informed,\\slides along walls} \\
\midrule
i & open field & 0 & 4 & 4.00 & 1.25 & 0.75 & 0.35 & 4.00 & 4.00 \\
ii & single detour & 1 & 3 & 3.00 & 0.70 & 0.70 & 0.40 & 3.00 & 3.00 \\
iii & two-wall slalom & 2 & 3 & 2.95 & 0.50 & 0.20 & 0.15 & 3.00 & 3.00 \\
iv & wide gate & 2 & 3 & 3.00 & 1.15 & 0.65 & 0.10 & 3.00 & 3.00 \\
v & one-sided walls & 2 & 3 & 3.00 & 0.40 & 0.15 & 0.05 & 3.00 & 3.00 \\
vi & long S-curve & 2 & 3 & 3.00 & 0.40 & 0.05 & 0.15 & 1.00 & 3.00 \\
vii & three-wall slalom & 3 & 4 & 3.95 & 0.45 & 0.30 & 0.40 & 4.00 & 4.00 \\
viii & four-wall slalom & 4 & 5 & 3.90 & 0.50 & 0.30 & 0.25 & 5.00 & 5.00 \\
ix & dense slalom & 4 & 4 & 3.45 & 0.45 & 0.35 & 0.35 & 3.00 & 4.00 \\
\midrule
& all nine & & 32 & 30.25 & 5.80 & 3.45 & 2.20 & 29.00 & 32.00 \\
\bottomrule
\end{tabular}}
\end{table}

\begin{table}[!htbp]\centering\small
\caption*{\textbf{Supplementary Table 2. The conditions of the forage arena.} Mean $\pm$ s.e.m. over the runs used: the first 20 completed runs of each condition in seed order, a rule fixed before any spare seed ran. Faulted, runs that ended in a physics fault, excluded from every number. None and All, the number of runs that reached no source and all three. Closest approach, the mean over the three sources. DNa02, spikes per 50 ms step summed over the pair. Wall contact, steps within 0.5 mm of the wall.}
\resizebox{\ifdim\width>\textwidth\textwidth\else\width\fi}{!}{%
\begin{tabular}{lrrrrrrrrrr}
\toprule
Condition & Runs & Faulted & \shortstack{Sources\\reached (of 3)} & None & All & \shortstack{Closest\\approach (mm)} & \shortstack{DNa02\\(spikes per step)} & \shortstack{Wall contact\\(steps)} & \shortstack{Path\\(mm)} & \shortstack{Leg motor\\rate (Hz)} \\
\midrule
Intact & 20 & 3 & 3.00 $\pm$ 0.00 & 0 & 20 & 1.15 $\pm$ 0.04 & 1.99 $\pm$ 0.02 & 0.0 $\pm$ 0.0 & 193 $\pm$ 5 & 154 $\pm$ 0 \\
Mushroom body lesioned & 20 & 0 & 0.00 $\pm$ 0.00 & 20 & 0 & 14.55 $\pm$ 0.69 & 0.81 $\pm$ 0.11 & 1.9 $\pm$ 1.9 & 220 $\pm$ 11 & 152 $\pm$ 0 \\
Rewired & 20 & 0 & 0.10 $\pm$ 0.07 & 18 & 0 & 11.00 $\pm$ 0.91 & 0.54 $\pm$ 0.15 & 1.9 $\pm$ 1.7 & 212 $\pm$ 11 & 98 $\pm$ 5 \\
Lesioned, valence pinned at 1.44 & 20 & 1 & 3.00 $\pm$ 0.00 & 0 & 20 & 1.08 $\pm$ 0.03 & 0.77 $\pm$ 0.03 & 0.0 $\pm$ 0.0 & 191 $\pm$ 5 & 154 $\pm$ 0 \\
Rewired, valence pinned at 1.44 & 20 & 1 & 2.55 $\pm$ 0.25 & 3 & 17 & 2.64 $\pm$ 0.96 & 0.46 $\pm$ 0.11 & 25.1 $\pm$ 15.6 & 186 $\pm$ 13 & 103 $\pm$ 6 \\
Lesioned, valence pinned at 0.7 & 20 & 0 & 3.00 $\pm$ 0.00 & 0 & 20 & 0.98 $\pm$ 0.07 & 0.74 $\pm$ 0.05 & 0.0 $\pm$ 0.0 & 190 $\pm$ 9 & 152 $\pm$ 0 \\
Lesioned, valence pinned at 0 & 20 & 0 & 0.00 $\pm$ 0.00 & 20 & 0 & 14.55 $\pm$ 0.69 & 0.81 $\pm$ 0.11 & 1.9 $\pm$ 1.9 & 220 $\pm$ 11 & 152 $\pm$ 0 \\
Intact, valence pinned at 0 & 20 & 0 & 0.00 $\pm$ 0.00 & 20 & 0 & 16.40 $\pm$ 0.19 & 2.08 $\pm$ 0.08 & 0.0 $\pm$ 0.0 & 226 $\pm$ 8 & 152 $\pm$ 0 \\
Random lesion of 4,608 neurons & 20 & 0 & 2.55 $\pm$ 0.25 & 3 & 17 & 3.04 $\pm$ 1.06 & 2.73 $\pm$ 0.33 & 0.0 $\pm$ 0.0 & 174 $\pm$ 10 & 131 $\pm$ 7 \\
Intact, looming signal at zero & 20 & 3 & 0.30 $\pm$ 0.21 & 18 & 2 & 10.72 $\pm$ 0.77 & 1.85 $\pm$ 0.01 & 233.1 $\pm$ 18.0 & 53 $\pm$ 8 & 152 $\pm$ 0 \\
\bottomrule
\end{tabular}}
\end{table}

\begin{table}[!htbp]\centering\small
\caption*{\textbf{Supplementary Table 3. Descending-neuron drive in the body at three receptor rates.} Net turn over the 2 s of drive (degrees; positive, to the left) and the change in speed from the settling window to the drive window (mm/s), mean $\pm$ s.e.m.; $n$, completed runs. The receptor neurons of both antennae are driven at the rate shown. Rewired, three permutations with seven seeds each at 120 Hz.}
\resizebox{\ifdim\width>\textwidth\textwidth\else\width\fi}{!}{%
\begin{tabular}{lrrrrrrrrrrrr}
\toprule
 & \multicolumn{3}{c}{0 Hz} & \multicolumn{3}{c}{40 Hz} & \multicolumn{3}{c}{120 Hz} & \multicolumn{3}{c}{120 Hz, rewired} \\
\cmidrule(lr){2-4}\cmidrule(lr){5-7}\cmidrule(lr){8-10}\cmidrule(lr){11-13}
Target & Turn & Speed & $n$ & Turn & Speed & $n$ & Turn & Speed & $n$ & Turn & Speed & $n$ \\
\midrule
DNa01, left & $-$219 $\pm$ 7 & 9.8 $\pm$ 0.1 & 20 & 86 $\pm$ 7 & 1.4 $\pm$ 0.1 & 20 & 116 $\pm$ 7 & 1.1 $\pm$ 0.1 & 20 & 55 $\pm$ 7 & 1.4 $\pm$ 0.2 & 21 \\
DNa01, right & $-$92 $\pm$ 13 & 4.6 $\pm$ 0.2 & 20 & $-$167 $\pm$ 5 & 0.0 $\pm$ 0.1 & 20 & $-$150 $\pm$ 7 & $-$0.4 $\pm$ 0.1 & 20 & 36 $\pm$ 9 & 2.0 $\pm$ 0.3 & 21 \\
DNa02, left & $-$269 $\pm$ 7 & 7.2 $\pm$ 0.1 & 20 & $-$1 $\pm$ 7 & 1.3 $\pm$ 0.1 & 20 & $-$16 $\pm$ 7 & 1.3 $\pm$ 0.1 & 20 & $-$7 $\pm$ 9 & 0.7 $\pm$ 0.1 & 21 \\
DNa02, right & 219 $\pm$ 7 & 4.3 $\pm$ 0.1 & 20 & $-$47 $\pm$ 6 & 0.4 $\pm$ 0.1 & 20 & $-$25 $\pm$ 6 & 0.2 $\pm$ 0.1 & 20 & $-$7 $\pm$ 9 & 1.7 $\pm$ 0.3 & 21 \\
DNa02, both & $-$85 $\pm$ 10 & 12.0 $\pm$ 0.1 & 20 & $-$48 $\pm$ 8 & 0.7 $\pm$ 0.1 & 20 & $-$38 $\pm$ 7 & 0.8 $\pm$ 0.1 & 20 & -- & -- & 0 \\
DNp09, left & 169 $\pm$ 8 & 1.7 $\pm$ 0.0 & 20 & 31 $\pm$ 7 & $-$2.1 $\pm$ 0.1 & 20 & 54 $\pm$ 6 & $-$2.2 $\pm$ 0.1 & 20 & 3 $\pm$ 6 & 0.9 $\pm$ 0.1 & 21 \\
Threat-response cluster, both & 7 $\pm$ 10 & 3.1 $\pm$ 0.0 & 20 & $-$114 $\pm$ 8 & $-$5.2 $\pm$ 0.1 & 20 & $-$45 $\pm$ 7 & $-$5.1 $\pm$ 0.1 & 20 & 5 $\pm$ 8 & 6.3 $\pm$ 0.2 & 21 \\
DNg12, left & $-$3 $\pm$ 5 & 2.3 $\pm$ 0.0 & 20 & 14 $\pm$ 6 & 2.4 $\pm$ 0.1 & 20 & 12 $\pm$ 5 & 1.8 $\pm$ 0.1 & 20 & 27 $\pm$ 11 & 2.2 $\pm$ 0.2 & 21 \\
DNp01, left & $-$96 $\pm$ 10 & 0.1 $\pm$ 0.0 & 20 & 16 $\pm$ 7 & 0.9 $\pm$ 0.1 & 20 & $-$16 $\pm$ 7 & 0.7 $\pm$ 0.1 & 20 & 9 $\pm$ 7 & 0.5 $\pm$ 0.1 & 21 \\
No drive & 7 $\pm$ 7 & $-$0.8 $\pm$ 0.0 & 20 & 21 $\pm$ 7 & 0.3 $\pm$ 0.1 & 20 & $-$7 $\pm$ 7 & 0.5 $\pm$ 0.1 & 20 & -- & -- & 0 \\
\bottomrule
\end{tabular}}
\end{table}

\begin{table}[!htbp]\centering\small
\caption*{\textbf{Supplementary Table 4. The odor code against the gain of the cholinergic local neurons.} The output synapses of the local neurons predicted cholinergic are scaled by the gain (1, the network as read). Active, the fraction of cells with at least one spike during a 1 s presentation, averaged over the two odors; overlap, the fraction shared between the two odors' 5\% most active cells; MBON spikes, the total over the presentation before training; depression, 1 minus the ratio of the MBON output after training on odor A (strength 0.5) to before. Means over the seeds shown.}
\resizebox{\ifdim\width>\textwidth\textwidth\else\width\fi}{!}{%
\begin{tabular}{rrrrrrrrrr}
\toprule
Gain & Seeds & \shortstack{Kenyon cells\\active} & \shortstack{Kenyon-cell\\overlap} & \shortstack{Projection\\neurons active} & \shortstack{Projection-\\neuron overlap} & \shortstack{MBON spikes,\\odor A} & \shortstack{MBON spikes,\\odor B} & \shortstack{Depression\\of A} & \shortstack{Depression\\of B} \\
\midrule
1 & 6 & 0.682 & 0.90 & 0.74 & 0.86 & 2368 & 2416 & 0.18 & 0.18 \\
0.5 & 6 & 0.431 & 0.76 & 0.54 & 0.57 & 1026 & 1182 & 0.18 & 0.15 \\
0.25 & 6 & 0.148 & 0.61 & 0.33 & 0.29 & 160 & 286 & 0.29 & 0.16 \\
0.15 & 6 & 0.061 & 0.36 & 0.21 & 0.11 & 52 & 111 & 0.00 & 0.00 \\
0.1 & 6 & 0.034 & 0.24 & 0.16 & 0.11 & 38 & 102 & $-$0.03 & 0.02 \\
0.075 & 6 & 0.027 & 0.26 & 0.15 & 0.09 & 32 & 96 & $-$0.02 & $-$0.01 \\
0.05 & 6 & 0.023 & 0.31 & 0.13 & 0.06 & 28 & 92 & 0.00 & 0.02 \\
0 & 6 & 0.018 & 0.41 & 0.10 & 0.09 & 24 & 82 & 0.04 & $-$0.02 \\
\midrule
\multicolumn{2}{l}{Every local neuron signed inhibitory} & 0.014 & & & & & & & \\
\bottomrule
\end{tabular}}
\end{table}

\begin{table}[!htbp]\centering\small
\caption*{\textbf{Supplementary Table 5. The sign of every MBON type under the balance readout.} The balance readout (Methods) reads the valence as the spike count of the approach-driving MBONs minus that of the avoidance-driving MBONs. A type drives avoidance when its transmitter is glutamate and approach when it is GABA or acetylcholine; the transmitter is the connectome table's verified entry where the type has one, else the published one, else the majority of the per-neuron predictions (references in the Methods). Compartments are the hemibrain names; the aversive compartments are six of the seven that the PPL1 dopamine neurons innervate (all but $\alpha'1$), the only ones the reinforcement reaches in the learning of Fig.~6. Cells are those the network simulates.}
\resizebox{\ifdim\width>\textwidth\textwidth\else\width\fi}{!}{%
\begin{tabular}{llrlllc}
\toprule
Type & Compartment & Cells & Transmitter & Transmitter source & Drives & \shortstack{Aversive\\compartment} \\
\midrule
MBON01 & $\gamma5\beta'2a$ & 2 & glutamate & verified in the table & avoidance &  \\
MBON02 & $\beta2\beta'2a$ & 2 & glutamate & verified in the table & avoidance &  \\
MBON03 & $\beta'2mp$ & 2 & glutamate & verified in the table & avoidance &  \\
MBON04 & $\beta'2mp$ bilateral & 2 & glutamate & verified in the table & avoidance &  \\
MBON05 & $\gamma4{>}\gamma1\gamma2$ & 2 & glutamate & verified in the table & avoidance &  \\
MBON06 & $\beta1{>}\alpha$ & 2 & glutamate & verified in the table & avoidance &  \\
MBON07 & $\alpha1$ & 4 & glutamate & verified in the table & avoidance &  \\
MBON09 & $\gamma3\beta'1$ & 4 & GABA & verified in the table & approach &  \\
MBON10 & $\beta'1$ & 12 & GABA & verified in the table & approach &  \\
MBON11 & $\gamma1$pedc${>}\alpha/\beta$ & 2 & GABA & verified in the table & approach & yes \\
MBON12 & $\gamma2\alpha'1$ & 4 & acetylcholine & verified in the table & approach & yes \\
MBON13 & $\alpha'2$ & 3 & acetylcholine & verified in the table & approach & yes \\
MBON14 & $\alpha3$ & 5 & acetylcholine & verified in the table & approach & yes \\
MBON15 & $\alpha'1$ & 4 & acetylcholine & verified in the table & approach &  \\
MBON15-like & $\alpha'1$ & 3 & acetylcholine & verified in the table & approach &  \\
MBON16 & $\alpha'3$ap & 2 & acetylcholine & verified in the table & approach & yes \\
MBON17 & $\alpha'3$m & 4 & acetylcholine & verified in the table & approach & yes \\
MBON17-like & $\alpha'3$m & 1 & acetylcholine & verified in the table & approach & yes \\
MBON18 & $\alpha2$sc & 2 & acetylcholine & verified in the table & approach & yes \\
MBON19 & $\alpha2$p3p & 6 & acetylcholine & verified in the table & approach & yes \\
MBON20 & $\gamma1\gamma2$ & 2 & GABA & verified in the table & approach & yes \\
MBON21 & $\gamma4\gamma5$ & 2 & acetylcholine & majority of the predictions & approach &  \\
MBON22 & calyx & 2 & acetylcholine & majority of the predictions & approach &  \\
MBON23 & $\alpha2$sp & 2 & acetylcholine & majority of the predictions & approach & yes \\
MBON24 & $\beta2\gamma5$ & 2 & -- & none & unsigned &  \\
MBON25, MBON34 & $\gamma1\gamma2$ / $\gamma2$ & 7 & glutamate & majority of the predictions & avoidance & yes \\
MBON26 & $\beta'2$d & 2 & acetylcholine & published (Methods) & approach &  \\
MBON27 & $\gamma5$d & 2 & acetylcholine & published (Methods) & approach &  \\
MBON28 & $\alpha'3$ & 2 & acetylcholine & verified in the table & approach & yes \\
MBON29 & $\gamma4\gamma5$ & 2 & -- & none & unsigned &  \\
MBON30 & $\gamma1\gamma2\gamma3$ & 2 & glutamate & published (Methods) & avoidance & yes \\
MBON31 & $\alpha'1$ & 3 & GABA & majority of the predictions & approach &  \\
MBON32 & $\gamma2$ & 2 & GABA & published (Methods) & approach & yes \\
MBON33 & $\gamma2\gamma3$ & 2 & acetylcholine & verified in the table & approach & yes \\
MBON35 & $\gamma2$ & 2 & acetylcholine & majority of the predictions & approach & yes \\
\midrule
\multicolumn{2}{l}{All types} & 104 & & & 75 approach, 25 avoidance, 4 unsigned & 50 cells \\
\bottomrule
\end{tabular}}
\end{table}

\begin{table}[!htbp]\centering\small
\caption*{\textbf{Supplementary Table 6. The in-body results under the two readouts of the valence.} The summed readout reads the valence as the summed spike count of the MBONs divided by 107 (81 in the matched variant); the balance readout, the one of the main text, reads it as the count of the approach-driving MBONs minus that of the avoidance-driving MBONs (Methods; Supplementary Table~5). Every in-body run that reads the valence was made under each readout (Supplementary Discussion). Runs with the valence pinned do not read it. Under the summed readout the reinforcement reached every plastic connection; under the balance readout it reaches the 8,377 connections of the aversive compartments, so the weight row, a mean over all 15,523, is not comparable between the columns. Means over 20 seeds (the first 20 completed runs in the forage arena).}
\resizebox{\ifdim\width>\textwidth\textwidth\else\width\fi}{!}{%
\begin{tabular}{p{9.2cm}ll}
\toprule
Measure & Summed readout & Balance readout \\
\midrule
\multicolumn{3}{l}{\textit{Nine arenas (Fig.~4): sources reached of 32 per fly (runs of 180 reaching none; reaching every source of the arena)}} \\
Intact & 29.20 (0; 159) & 30.25 (0; 161) \\
Rewired & 3.50 (115; 0) & 5.80 (92; 2) \\
Mushroom body lesioned & 3.45 (117; 0) & 3.45 (117; 0) \\
Correlated random walk at the fly's speed & 2.05 & 2.20 \\
\midrule
\multicolumn{3}{l}{\textit{Forage arena (Fig.~5 b to d): sources reached of 3 (runs of 20 reaching none; reaching all three)}} \\
Intact & 2.40 (4; 16) & 3.00 (0; 20) \\
Mushroom body lesioned & 0.00 (20; 0) & 0.00 (20; 0) \\
Rewired & 0.15 (18; 0) & 0.10 (18; 0) \\
Lesioned, valence pinned at 1.44 & 3.00 (0; 20) & 3.00 (0; 20) \\
Rewired, valence pinned at 1.44 & 2.55 (3; 17) & 2.55 (3; 17) \\
Lesioned, valence pinned at 0.7 & 3.00 (0; 20) & 3.00 (0; 20) \\
Lesioned, valence pinned at 0 & 0.00 (20; 0) & 0.00 (20; 0) \\
Intact, valence pinned at 0 & 0.00 (20; 0) & 0.00 (20; 0) \\
Random lesion of 4,608 neurons & 2.85 (1; 19) & 2.55 (3; 17) \\
Intact, looming signal set to zero & 0.15 (19; 1) & 0.30 (18; 2) \\
\midrule
\multicolumn{3}{l}{\textit{Motor replay (Fig.~5 e, g): fraction of sources per fly}} \\
Closed loop, replay & 0.940, 0.030 & 0.940, 0.020 \\
Exact Wilcoxon $W$, $P$; flies with closed loop above replay & 0, $1.9\times10^{-6}$; 20 of 20 & 0, $1.9\times10^{-6}$; 20 of 20 \\
\midrule
\multicolumn{3}{l}{\textit{Graded removal (Fig.~5 h, i): sources reached of 3}} \\
20\% at random & 2.47 & 2.50 \\
50\% at random & 0.26 & 0.30 \\
1\% by traffic & 0.00 & 0.00 \\
\midrule
\multicolumn{3}{l}{\textit{Learning, the network as read (Fig.~6 b, e to h and j)}} \\
Weight of all Kenyon-cell-to-MBON connections, mV, naive $\to$ after pairing & 0.500 $\to$ 0.128 & 0.500 $\to$ 0.298 \\
Valence, naive $\to$ after (range after): odor + dopamine & 1.44 $\to$ 0.24 (0.24 to 0.25) & 1.41 $\to$ $-$0.33 ($-$0.34 to $-$0.32) \\
Valence, naive $\to$ after (range after): dopamine alone & 1.44 $\to$ 0.29 (0.28 to 0.29) & 1.41 $\to$ $-$0.29 ($-$0.30 to $-$0.28) \\
Valence, naive $\to$ after (range after): no plasticity & 1.44 $\to$ 1.44 (1.44 to 1.45) & 1.41 $\to$ 1.41 (1.40 to 1.42) \\
Flies reaching the source, naive, after: odor + dopamine & 20 of 20, 19 of 20 & 20 of 20, 0 of 20 \\
Flies reaching the source, naive, after: dopamine alone & 20 of 20, 20 of 20 & 20 of 20, 0 of 20 \\
Flies reaching the source, naive, after: no plasticity & 20 of 20, 20 of 20 & 20 of 20, 20 of 20 \\
\midrule
\multicolumn{3}{l}{\textit{Learning, the matched variant (Fig.~6 i, k)}} \\
Weight of all Kenyon-cell-to-MBON connections, mV, naive $\to$ after pairing & 0.500 $\to$ 0.182 & 0.500 $\to$ 0.328 \\
Valence, naive $\to$ after (range after): odor + dopamine & 1.44 $\to$ 0.25 (0.25 to 0.26) & 1.40 $\to$ $-$0.53 ($-$0.54 to $-$0.52) \\
Valence, naive $\to$ after (range after): dopamine alone & 1.44 $\to$ 1.42 (1.41 to 1.42) & 1.40 $\to$ 1.38 (1.36 to 1.40) \\
Valence, naive $\to$ after (range after): no plasticity & 1.44 $\to$ 1.44 (1.43 to 1.44) & 1.40 $\to$ 1.40 (1.38 to 1.42) \\
Flies reaching the source, naive, after: odor + dopamine & 20 of 20, 19 of 20 & 20 of 20, 0 of 20 \\
Flies reaching the source, naive, after: dopamine alone & 20 of 20, 20 of 20 & 20 of 20, 20 of 20 \\
Flies reaching the source, naive, after: no plasticity & 20 of 20, 20 of 20 & 20 of 20, 20 of 20 \\
\bottomrule
\end{tabular}}
\end{table}

\begin{table}[!htbp]\centering\small
\caption*{\textbf{Supplementary Table 7. The learning experiment under the four combinations of valence readout and reinforcement.} The four combinations of Supplementary Fig.~4 in both models: the valence read as the summed MBON count or as the balance of approach-driving and avoidance-driving MBONs, with the reinforcement reaching every Kenyon-cell-to-MBON connection (15,523) or those of the aversive compartments (8,377). The first column is the summed readout with the reinforcement in every compartment (Supplementary Discussion) and the last the analysis of the main text (Fig.~6); the two between them change one of the two at a time, their predictions written down before they ran. Each condition has 20 flies. Valence, the mean over the post-test episode. Closest approach, to the source in the post test; a fly reaches the source when it comes within 4 mm. Tests, two-sided Mann--Whitney $U$ on the closest approach, with Cliff's $\delta$ and its 95\% bootstrap interval where the groups overlap; no correction for multiple comparisons. One of the 16 tests has $P$ between 0.001 and 0.05: the matched variant with the sum readout and the reinforcement in the aversive compartments, odor + dopamine against dopamine alone ($P = 0.014$; medians 0.27 and 0.18 mm, every fly of both groups reaching the source). Without plasticity the reach of the reinforcement has no effect, and the two no-plasticity columns of a readout are the same simulation run twice: they agree fly for fly.}
\resizebox{\ifdim\width>\textwidth\textwidth\else\width\fi}{!}{%
\begin{tabular}{lllll}
\toprule
 & \multicolumn{2}{c}{Summed readout, reinforcement in} & \multicolumn{2}{c}{Balance readout, reinforcement in} \\
\cmidrule(lr){2-3}\cmidrule(lr){4-5}
Measure & \shortstack{every\\compartment} & \shortstack{the aversive\\compartments} & \shortstack{every\\compartment} & \shortstack{the aversive\\compartments} \\
\midrule
\multicolumn{5}{l}{\textit{The network as read}} \\
Valence after training (range over flies) & & & & \\
\quad odor + dopamine & 0.24 (0.24 to 0.25) & 0.83 (0.82 to 0.83) & 0.03 (0.03 to 0.04) & $-$0.33 ($-$0.34 to $-$0.32) \\
\quad dopamine alone & 0.29 (0.28 to 0.29) & 0.83 (0.83 to 0.84) & 0.03 (0.01 to 0.08) & $-$0.29 ($-$0.30 to $-$0.28) \\
\quad no plasticity & 1.44 (1.44 to 1.45) & 1.44 (1.44 to 1.45) & 1.41 (1.40 to 1.42) & 1.41 (1.40 to 1.42) \\
Flies reaching the source after training, of 20 & & & & \\
\quad odor + dopamine & 19 & 20 & 2 & 0 \\
\quad dopamine alone & 20 & 20 & 1 & 0 \\
\quad no plasticity & 20 & 20 & 20 & 20 \\
Closest approach after training, median (range), mm & & & & \\
\quad odor + dopamine & 0.23 (0.08 to 4.92) & 0.26 (0.08 to 1.07) & 11.25 (2.73 to 14.25) & 19.56 (18.06 to 21.02) \\
\quad dopamine alone & 0.18 (0.07 to 1.97) & 0.28 (0.03 to 1.25) & 9.98 (2.76 to 15.40) & 19.54 (17.61 to 20.48) \\
\quad no plasticity & 0.19 (0.07 to 1.16) & 0.19 (0.07 to 1.16) & 0.15 (0.09 to 0.40) & 0.15 (0.09 to 0.40) \\
Closest approach, odor + dopamine against no plasticity & & & & \\
\quad $P$ & 0.23 & 0.13 & $6.8\times10^{-8}$ & $6.8\times10^{-8}$ \\
\quad Cliff's $\delta$ (95\% interval) & $+$0.23 ($-$0.14 to $+$0.56) & $+$0.28 ($-$0.08 to $+$0.62) & $+$1.00 & $+$1.00 \\
Closest approach, odor + dopamine against dopamine alone & & & & \\
\quad $P$ & 0.30 & 0.69 & 0.39 & 0.49 \\
\quad Cliff's $\delta$ (95\% interval) & $+$0.20 ($-$0.17 to $+$0.54) & $-$0.07 ($-$0.45 to $+$0.29) & $+$0.16 ($-$0.21 to $+$0.52) & $+$0.13 ($-$0.25 to $+$0.49) \\
\midrule
\multicolumn{5}{l}{\textit{The matched variant}} \\
Valence after training (range over flies) & & & & \\
\quad odor + dopamine & 0.25 (0.25 to 0.26) & 0.84 (0.83 to 0.84) & $-$0.02 ($-$0.03 to 0.00) & $-$0.53 ($-$0.54 to $-$0.52) \\
\quad dopamine alone & 1.42 (1.41 to 1.42) & 1.43 (1.42 to 1.43) & 1.39 (1.37 to 1.41) & 1.38 (1.36 to 1.40) \\
\quad no plasticity & 1.44 (1.43 to 1.44) & 1.44 (1.43 to 1.44) & 1.40 (1.38 to 1.42) & 1.40 (1.38 to 1.42) \\
Flies reaching the source after training, of 20 & & & & \\
\quad odor + dopamine & 19 & 20 & 1 & 0 \\
\quad dopamine alone & 20 & 20 & 20 & 20 \\
\quad no plasticity & 20 & 20 & 20 & 20 \\
Closest approach after training, median (range), mm & & & & \\
\quad odor + dopamine & 0.29 (0.10 to 4.28) & 0.27 (0.00 to 1.20) & 10.18 (2.06 to 15.77) & 21.08 (20.39 to 21.60) \\
\quad dopamine alone & 0.22 (0.01 to 1.30) & 0.18 (0.03 to 0.89) & 0.21 (0.05 to 1.18) & 0.27 (0.01 to 1.18) \\
\quad no plasticity & 0.26 (0.06 to 1.17) & 0.26 (0.06 to 1.17) & 0.23 (0.02 to 1.24) & 0.23 (0.02 to 1.24) \\
Closest approach, odor + dopamine against no plasticity & & & & \\
\quad $P$ & 0.31 & 0.58 & $6.8\times10^{-8}$ & $6.8\times10^{-8}$ \\
\quad Cliff's $\delta$ (95\% interval) & $+$0.19 ($-$0.18 to $+$0.52) & $+$0.10 ($-$0.25 to $+$0.46) & $+$1.00 & $+$1.00 \\
Closest approach, odor + dopamine against dopamine alone & & & & \\
\quad $P$ & 0.21 & 0.014 & $6.8\times10^{-8}$ & $6.8\times10^{-8}$ \\
\quad Cliff's $\delta$ (95\% interval) & $+$0.24 ($-$0.12 to $+$0.57) & $+$0.46 ($+$0.12 to $+$0.75) & $+$1.00 & $+$1.00 \\
\bottomrule
\end{tabular}}
\end{table}

\begin{table}[!htbp]\centering\small
\caption*{\textbf{Supplementary Table 8. The 24 groups of Fig.~1 and their defining labels.} The groups that Fig.~1 b to f read from, with the number of simulated neurons in each. A neuron takes the first of the 13 coarse rules it matches, in this order: visual projection, olfactory, mushroom body, lateral horn, central complex, descending, ascending, leg motor, other motor, other sensory, optic lobe, VNC intrinsic, other central (the rules read the neuron table's super\_class, cell\_class, cell\_sub\_class, cell\_type and region columns; the names with underscores are the table's own labels). The central-complex groups split the central-complex rule (cell\_class containing central\_complex) by cell\_type prefix, the sensory groups split the other-sensory rule (super\_class sensory, sensory\_ascending or sensory\_descending, not olfactory receptor or photoreceptor) by cell\_class, and the leg motor groups split cell\_class leg\_motor\_neuron by the segment named in cell\_sub\_class. Non-neuronal units (glia, trachea) count where their labels place them.}
\resizebox{\ifdim\width>\textwidth\textwidth\else\width\fi}{!}{%
\begin{tabular}{p{3.6cm}p{10.4cm}r}
\toprule
Group & Defining labels & Neurons \\
\midrule
optic lobe & region optic\_lobe & 93,925 \\
visual projection & super\_class visual\_projection & 7,284 \\
olfactory & cell\_class olfactory\_receptor\_neuron & 2,978 \\
mushroom body & cell\_class kenyon\_cell or mushroom\_body\_output\_neuron & 4,608 \\
lateral horn & cell\_class containing lateral\_horn & 1,655 \\
other central & super\_class central\_brain\_intrinsic not matched by an earlier rule (the antennal-lobe projection and local neurons among them), and the 8,409 neurons matching no rule & 30,575 \\
ellipsoid body & a split of central complex: cell\_type EPG, PEG, PEN, EL, ER or EB & 397 \\
fan-shaped body & a split of central complex: cell\_type FB, FC, FR, FS, PFL, PFN, PFR, PFG, vDelta or hDelta & 2,362 \\
noduli & a split of central complex: cell\_type NO or LNO & 6 \\
protocerebral bridge & a split of central complex: cell\_type Delta7, IbSpsP or PB & 68 \\
central complex, other & a split of central complex: the other central-complex cell types & 98 \\
bristle (touch) & a split of other sensory: cell\_class bristle\_neuron & 6,017 \\
campaniform & a split of other sensory: cell\_class campaniform\_sensillum\_neuron & 575 \\
chordotonal & a split of other sensory: cell\_class chordotonal\_organ\_neuron & 2,115 \\
gustatory & a split of other sensory: cell\_class taste\_bristle\_gustatory\_neuron or taste\_peg\_gustatory\_neuron & 1,362 \\
hair plate & a split of other sensory: cell\_class hair\_plate\_neuron & 320 \\
sensory, other & a split of other sensory: the other sensory cell classes & 1,275 \\
VNC intrinsic & super\_class ventral\_nerve\_cord\_intrinsic & 12,838 \\
descending & super\_class descending & 1,316 \\
ascending & super\_class ascending or ascending\_visceral\_circulatory & 1,854 \\
front leg motor & a split of leg motor: cell\_sub\_class naming the front leg & 139 \\
middle leg motor & a split of leg motor: cell\_sub\_class naming the middle leg & 126 \\
hind leg motor & a split of leg motor: cell\_sub\_class naming the hind leg & 126 \\
other motor & super\_class motor, not leg\_motor\_neuron (and leg motor neurons without a segment) & 414 \\
\midrule
all & & 172,433 \\
\bottomrule
\end{tabular}}
\end{table}

\begin{table}[!htbp]\centering\small
\caption*{\textbf{Supplementary Table 9. The learning experiment with and without the MBON's spikes in the plasticity rule.} In the analysis of the main text (Fig.~6) a plastic connection accumulates a depression of $\eta$ times its eligibility times its MBON's spikes in each step with the reinforcement on. The control left out the MBON's spikes, a factor that the depression in the fly does not need (Methods). It ran the odor + dopamine and dopamine-alone conditions again in both models with the same seeds, and its predictions were written down before it ran. Each condition has 20 flies, and each control run is identical to the reported run of its seed until training begins. Weight, the mean over the 8,377 plastic connections (0.513 mV before training). Valence, the mean over the post-test episode. Closest approach, to the source in the post test; a fly reaches the source when it comes within 4 mm. Devaluation, the fall from the naive to the post test in the fraction of the start distance closed. Tests, two-sided Mann--Whitney $U$, with Cliff's $\delta$ and its 95\% bootstrap interval where the groups overlap; no correction for multiple comparisons. The registered tests are those of the Methods; the depression counts as expressed only if the flies given dopamine alone still reach the source. The runs without plasticity do not use the rule, and those of Fig.~6 serve both columns of a model (all 20 of them reach the source in each model).}
\resizebox{\ifdim\width>\textwidth\textwidth\else\width\fi}{!}{%
\begin{tabular}{lllll}
\toprule
 & \multicolumn{2}{c}{The network as read} & \multicolumn{2}{c}{The matched variant} \\
\cmidrule(lr){2-3}\cmidrule(lr){4-5}
Measure & \shortstack{with the factor\\(Fig.~6)} & \shortstack{without\\the factor} & \shortstack{with the factor\\(Fig.~6)} & \shortstack{without\\the factor} \\
\midrule
Weight of the plastic connections after training, mV & & & & \\
\quad odor + dopamine & 0.138 & 0.089 & 0.195 & 0.139 \\
\quad dopamine alone & 0.183 & 0.128 & 0.508 & 0.455 \\
Valence after training (range over flies) & & & & \\
\quad odor + dopamine & $-$0.33 ($-$0.34 to $-$0.32) & $-$0.32 ($-$0.33 to $-$0.32) & $-$0.53 ($-$0.54 to $-$0.52) & $-$0.53 ($-$0.54 to $-$0.52) \\
\quad dopamine alone & $-$0.29 ($-$0.30 to $-$0.28) & $-$0.30 ($-$0.31 to $-$0.29) & 1.38 (1.36 to 1.40) & 0.90 (0.87 to 0.92) \\
Flies reaching the source after training, of 20 & & & & \\
\quad odor + dopamine & 0 & 0 & 0 & 0 \\
\quad dopamine alone & 0 & 0 & 20 & 20 \\
Closest approach after training, median (range), mm & & & & \\
\quad odor + dopamine & 19.56 (18.06 to 21.02) & 19.79 (18.64 to 20.74) & 21.08 (20.39 to 21.60) & 20.89 (20.18 to 21.78) \\
\quad dopamine alone & 19.54 (17.61 to 20.48) & 19.78 (17.51 to 20.80) & 0.27 (0.01 to 1.18) & 0.23 (0.06 to 1.22) \\
Devaluation, median (range) & & & & \\
\quad odor + dopamine & 0.79 (0.72 to 0.85) & 0.79 (0.73 to 0.83) & 0.85 (0.80 to 0.88) & 0.84 (0.79 to 0.88) \\
\quad dopamine alone & 0.78 (0.71 to 0.82) & 0.78 (0.70 to 0.84) & 0.00 ($-$0.05 to 0.04) & 0.00 ($-$0.04 to 0.04) \\
Closest approach, odor + dopamine against dopamine alone & & & & \\
\quad $P$ & 0.49 & 0.88 & $6.8\times10^{-8}$ & $6.8\times10^{-8}$ \\
\quad Cliff's $\delta$ (95\% interval) & $+$0.13 ($-$0.25 to $+$0.49) & $+$0.03 ($-$0.32 to $+$0.38) & $+$1.00 & $+$1.00 \\
Devaluation, odor + dopamine against dopamine alone & & & & \\
\quad $P$ & 0.36 & 0.68 & $6.8\times10^{-8}$ & $6.8\times10^{-8}$ \\
\quad Cliff's $\delta$ (95\% interval) & $+$0.17 ($-$0.21 to $+$0.53) & $+$0.08 ($-$0.28 to $+$0.44) & $+$1.00 & $+$1.00 \\
Registered tests (Methods) & & & & \\
\quad associative & no & no & yes & yes \\
\quad expressed & no & no & yes & yes \\
\bottomrule
\end{tabular}}
\end{table}

\restoregeometry
\end{document}